\documentclass[aps,natphysics,superscriptaddress,two column,balance,preprintnumbers,pagewise,notitlepage,bibnotes]{revtex4-2}
\usepackage{latexsym}
\usepackage{subcaption}
\usepackage{balance}
\usepackage{caption}
\usepackage{float}
\usepackage[margin=0.6in]{geometry}
\usepackage[dvips]{color}
\usepackage[pdftex]{graphicx}
\usepackage{amsmath}
\usepackage{amssymb}
\usepackage{esdiff}
\usepackage{ragged2e}
\usepackage{lineno}
\usepackage{gensymb}
\usepackage{enumerate}
\usepackage{hyperref}
\usepackage{mathrsfs}
\usepackage[usenames,dvipsnames]{xcolor}
\usepackage{float}
\begin{document}
 
\title{\textbf{\large Emergent phases in graphene flat bands}}
\author{Saisab Bhowmik}
\email{saisabb@iisc.ac.in}
\affiliation{\textit{Department of Instrumentation and Applied Physics, Indian Institute of Science, Bangalore, 560012, India}}

\author{Arindam Ghosh}
\affiliation{\textit{Department of Physics, Indian Institute of Science, Bangalore, 560012, India}}
\affiliation{\textit{Centre for Nano Science and Engineering, Indian Institute of Science, Bangalore 560 012, India}}

\author{U. Chandni}
\email{chandniu@iisc.ac.in}
\affiliation{\textit{Department of Instrumentation and Applied Physics, Indian Institute of Science, Bangalore, 560012, India}}

\pacs{}
\begin{abstract}
\textbf{Electronic correlations in two-dimensional materials play a crucial role in stabilising emergent phases of matter.~The realisation of correlation-driven phenomena in graphene has remained a longstanding goal, primarily due to the absence of strong electron-electron interactions within its low-energy bands.~In this context, magic-angle twisted bilayer graphene has recently emerged as a novel platform featuring correlated phases favoured by the low-energy flat bands of the underlying moir\'{e} superlattice.~Notably, the observation of correlated insulators and superconductivity has garnered significant attention, leading to substantial progress in theoretical and experimental studies aiming to elucidate the origin and interplay between these two phases.~A wealth of correlated phases with unprecedented tunability was discovered subsequently, including orbital ferromagnetism, Chern insulators, strange metallicity, density waves, and nematicity.~However, a comprehensive understanding of these closely competing phases remains elusive.~The ability to controllably twist and stack multiple graphene layers has enabled the creation of a whole new family of moir\'{e} superlattices with myriad properties being discovered at a fast pace. Here, we review the progress and development achieved so far, encompassing the rich phase diagrams offered by these graphene-based moir\'{e} systems.~Additionally, we discuss multiple phases recently observed in non-moir\'{e} multilayer graphene systems.~Finally, we outline future opportunities and challenges for the exploration of hidden phases in this new generation of moir\'{e} materials.}
\end{abstract}

\maketitle

\section{Introduction}

Strong electronic correlations form one of the cornerstones of condensed matter physics, laying the foundations for our understanding of many exotic phases of matter. At the simplest level of understanding, the quantum mechanical description and motion of an electron in `correlated materials' are inherently coupled to the behaviour of the other electrons, whereby traditional single-particle band structure calculations fail to explain their properties. A diverse range of many-body phases such as high $T_c$ superconductivity~\cite{da2014ubiquitous, anderson2004physics, cyrot1992introduction}, fractional quantum Hall effect~\cite{RevModPhys.71.298, PhysRevB.41.7653, bolotin2009observation, eisenstein1990fractional}, nematic ordering~\cite{PhysRevLett.82.394, ishida2020novel, PhysRevMaterials.2.114601}, charge density waves~\cite{chen2016charge}, and non-Fermi liquid behaviour~\cite{PhysRevB.81.184519, RevModPhys.73.797, PhysRevB.97.085118} have been discovered, holding promises for both fundamental science and technology.~Despite years of systematic research, many of these materials and their phase diagrams are not fully understood, partly due to insufficient control parameters to tune the overall experimental space, and partly due to the complexity of the theoretical calculations that incorporate strong Coulomb repulsions. While we make progress in better understanding the existing family of correlated systems, the search for new systems capable of addressing various longstanding roadblocks in both theory and experiments must continue.\\

Two-dimensional (2D) layered materials form a potential platform to explore correlations and resultant exotic phases of matter. The field has rapidly expanded since the isolation of monolayer graphene from bulk graphite~\cite{novoselov2004electric}, with many new candidates that are conductors, insulators, semi-metals, or semiconductors.~The strong van der Waals interactions between these layers enable us to stack them vertically to form a hybrid called a `van der Waals heterostructure'~\cite{novoselov20162d}, which can show emergent phenomena otherwise absent in the single-layer building blocks. These novel hybrids provide versatility to mitigate some of the experimental challenges encountered in conventional 2D electron systems, such as heterostructures of GaAs/AlGaAs and InGaAsP/InP.~Unlike other nanostructures, the absence of dangling bonds and atomic trap states leads to extraordinary electrical and optical properties in 2D layered materials.~Although efforts have been devoted to making interfaces including 2D-3D, 2D-1D, and 2D-0D, the archetypal 2D-2D heterostructures~\cite{jariwala2017mixed}, in particular, have shown immense potential for engineering novel devices. However, realisation of correlated physics in van der Waals materials has been a difficult task, primarily due to the inherent weak electron-electron interactions. An effective approach to induce strong electronic correlations involves the design of flat electronic bands with a high density of states (DOS), amplifying the ratio of Coulomb repulsion to kinetic energy.~A flat band is, therefore, an ideal platform to investigate correlated phases since electrons within a small energy window can exhibit collective behaviour, beyond the independent electron approximation.~The band flattening also reduces the Fermi velocity, leading to the localisation of electrons and the emergence of insulating states.~The scheme is akin to the well-known quantum Hall effect~\cite{cage2012quantum, PhysRevB.41.7653}, where the energy dispersion of a 2D electron gas undergoes significant modifications in the presence of a perpendicular magnetic field, resulting in flat and discrete energy levels. 
\begin{figure*}[bth]
\includegraphics[width=1.0\textwidth]{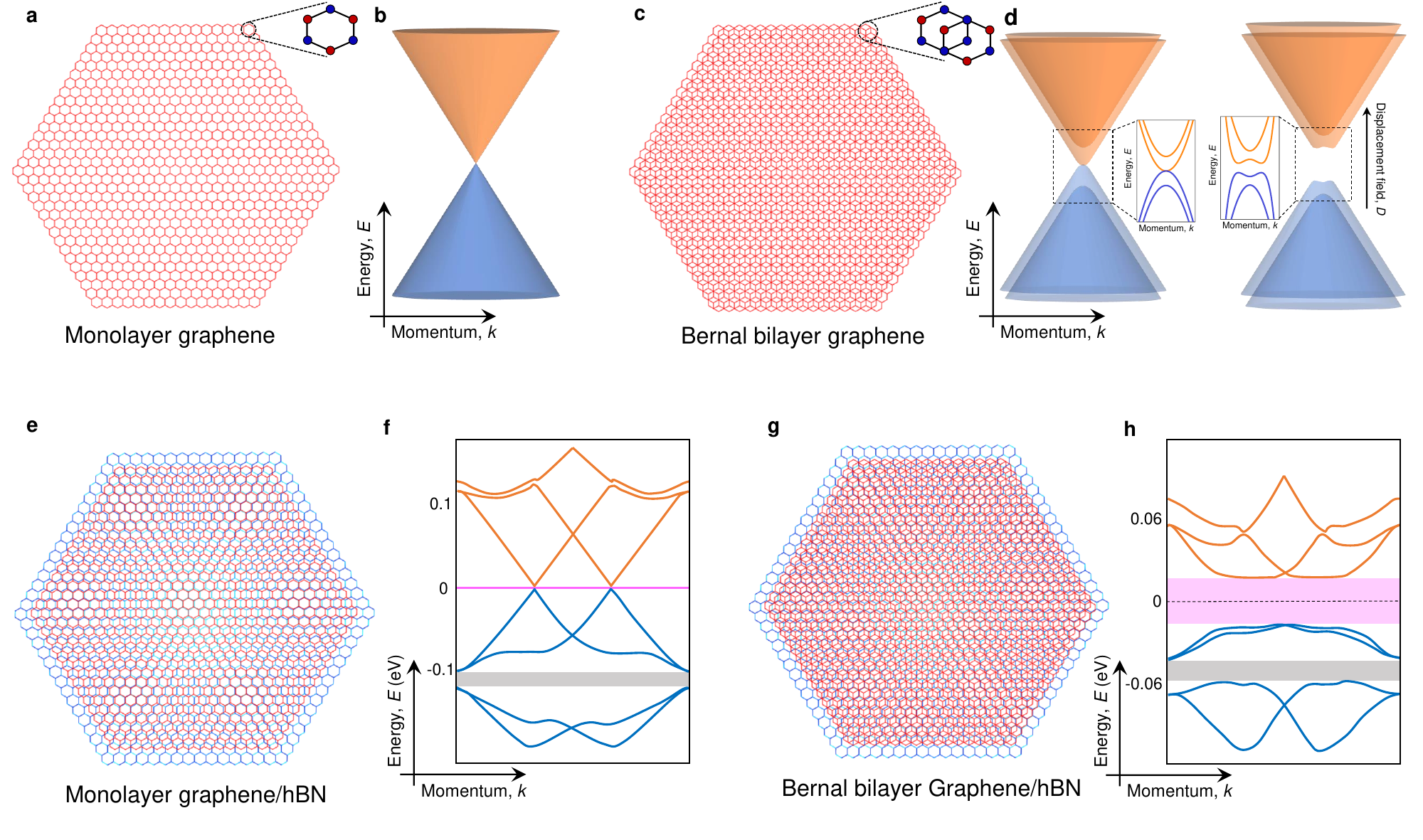}
\captionsetup{justification=raggedright,singlelinecheck=false}
\justify{\textbf{Fig.~1.~Lattice and electronic band structures of various graphene systems}.~\textbf{a.}~Monolayer graphene consists of one layer of carbon atoms arranged in a honeycomb lattice.~The different atomic environments at the two adjacent corners of the honeycomb lattice requires two atoms A (red) and B (blue) to define the unit cell.~\textbf{b.}~The low-energy band dispersion around the corners of the Brillouin zone ($K$ and $K^\prime$) is linear with the conduction (orange) and valence (blue) bands touching at the Dirac point.~\textbf{c.}~Bernal bilayer graphene consists of two monolayer graphene layers.~The most stable configuration in terms of structural relaxation is the AB stacking order where A sublattice of the top layer sits just on B sublattice of the bottom layer.~\textbf{d.}~In BLG, the total number of bands around $K$ and $K^\prime$ points becomes 4 (2 conduction, 2 valence bands), which are parabolic.~The application of an out-of-plane electric field opens up a gap at the parabolic band touching by breaking the two-fold rotational $C_2$ symmetry. Here, we have presented the simplest parabolic bands, however, the addition of the skew-interlayer hopping term can modify the low-energy band structure significantly by producing four Fermi surface pockets near the CNP. Such a phenomenon is known as `trigonal warping' and has been realised in experiments on ultra-clean BLG samples~\cite{mayorov2011interaction, jayaraman2021evidence}.~\textbf{e.}~Illustration of the moir\'{e} pattern formed when monolayer graphene is aligned with hBN.~\textbf{f.} Recreated schematic of the band structure obtained for monolayer graphene/hBN using the tight-binding model in ref.~\cite{PhysRevB.90.155406}. A bandgap of $\sim20$~meV between the first and second band opens on the hole side~(shaded by gray colour). At the CNP, there is a tiny gap of about 2 meV~(shaded by pink colour).~\textbf{g.}~Moir\'{e} pattern of Bernal BLG aligned with hBN.~\textbf{h.}~Recreated schematic of the band structure calculated by the tight-binding model in ref.~\cite{PhysRevB.90.155406}. At the CNP, a relatively large energy gap of about 40 meV in comparison to monolayer graphene/hBN is observed. However, the magnitude of the band gap between the first and second valence band remains similar to that of monolayer graphene/hBN.}
\end{figure*}
In recent years, advancements in layer-by-layer heterostructure assembly have enabled the manipulation and control of band structures, opening new avenues to tailor the interactions and making van der Waals hybrids the most sought-after correlated platform being investigated currently.\\

This article focuses on the recent developments in engineering flat bands in graphene-based systems and the efforts to uncover various correlated phases. Remarkably, when two graphene layers are misoriented to a specific angle, known as the `magic' angle~\cite{bistritzer2011moire}, an unexpected profusion of correlations arises, giving rise to various properties ranging from strongly insulating states, superconductivity to magnetism.~These phases exhibit high tunability under different experimental conditions, with some of them previously predicted by theoretical models, while most others being surprising new discoveries. The presence of an exotic phase diagram poses several critical questions about the nature of the ground states, prompting extensive investigations as well as theoretical modeling of the underlying mechanisms. The field has seen a steep rise in interest following the discovery of insulating and superconducting phases in magic-angle twisted bilayer graphene (MATBG) in 2018~\cite{cao2018correlated, cao2018unconventional}. While the primary focus of this review will be on the extensive data available for MATBG, we will also emphasize the broad possibilities offered by other twisted graphene systems, as well as Bernal and rhombohedral graphene.\\

The review is structured as follows. We begin by providing the minimal theoretical background necessary to understand monolayer and bilayer graphene systems, followed by a brief introduction to moir\'{e} superlattices, including a general classification of twisted bilayer graphene (TBG) and twisted multilayer graphene. We will then discuss various emergent properties in a pedagogical style, addressing the current state of understanding, as well as open challenges. This section is written in a comprehensive manner covering correlated insulators, superconductivity, flavour symmetries, and topological phases. A  discussion on scanning and imaging techniques follows, which provides a broad overview beyond transport studies. We will then briefly focus on the recent developments in Bernal and rhombohedral graphene systems.  In the end, we discuss the future opportunities to design a whole new class of quantum devices and the challenges for exploring hidden phases in these new-generation quantum materials. For previous reviews on TBG, readers may refer to~\cite{andrei2020graphene, balents2020superconductivity}. For twisted transition metal dichalcogenides (TMDC) and other semiconductor moir\'{e} materials, readers may refer to the recent review by Mak and Shan~\cite{mak2022semiconductor}. Lastly, readers who are not familiar with the basics of monolayer and bilayer graphene may find it useful to consult~\cite{RevModPhys.81.109, mccann2013electronic}.

\section{Electronic structure of monolayer and Bernal bilayer graphene}

Graphene has been the subject of many experimental and theoretical studies since its experimental discovery in 2004~\cite{novoselov2004electric, novoselov2007room, bolotin2009observation, PhysRevLett.95.146801, PhysRevLett.95.226801, roy2013graphene}.~We start with some preliminary concepts in monolayer graphene, which are relevant to understanding the intriguing emergent phases discussed in this article.~The carbon atoms in monolayer graphene are arranged in a honeycomb structure, and the Bravais lattice is triangular with a basis of two atoms per unit cell, denoted as A and B sublattices~(Fig.~1a).~The first Brillouin zone in the reciprocal space is hexagonal with the corners denoted as $K$ and $K^\prime$ which are also known as Dirac points.~In the tight-binding model, the Hamiltonian is constructed considering electrons hopping to both nearest- and next nearest-neighbours.~The low-energy band diagram obtained by expanding the energy eigenvalues around $K$ and $K^\prime$ is of the form, $E_\pm \approx \pm v_F|\textbf{q}| + O [(q/K)^2]$, where $\textbf{q}$ is the momentum measured relative to $K$ and $K^\prime$ points, $a=0.246$~nm is the lattice constant of graphene, and $v_F=3ta/2$ is the Fermi velocity with $t$ being the nearest-neighbour hopping energy~\cite{RevModPhys.81.109}.~The linear bandstructure (Fig.~1b) gives two unconventional characteristics to the charge carriers in graphene:~(1)~The motion of electrons can be quantum mechanically described by a massless Dirac equation.~(2)~The DOS per unit cell $\rho(E)=\frac{2A_c}{\pi}\frac{|E|}{v_F^2}$ varies linearly in energy $E$, where $A_c$ is the area of the unit cell given by $3\sqrt{3}a^2/2$.~Four fold degeneracy which is typical of all graphene-based systems discussed in this article arises from the two valley ($K$, $K^\prime$) and two spin ($\uparrow, \downarrow$) degrees of freedom.~The Dirac points at $K$ and $K^\prime$ points are protected by a composite of $C_2\mathcal{T}$ symmetry where $C_{2}$ is the two-fold rotational and $\mathcal{T}$ is the time-reversal symmetry.~In experiments, the band degeneracy can be lifted by aligning graphene with hexagonal boron nitride (hBN) that breaks the $C_2$ symmetry, or applying a magnetic field that breaks the $\mathcal{T}$ symmetry, both of which will be discussed in detail later in this article.\\

Next we discuss the electronic structure of Bernal bilayer graphene (BLG), which consists of two coupled graphene monolayers~\cite{mccann2013electronic}.~There are four atoms in the unit cell, labelled as A1, B1 on the bottom layer and A2, B2 on the top layer.~The layers are arranged in such a way that one of the atoms from the top layer A2 (B2) is directly above an atom, B1 (A1) from the bottom layer and thus the overall structure is seen as AB (BA) stacking~(Fig.~1c).~The overlapping atomic sites from the two different layers within the unit cell are referred as `dimer' sites and the two atoms that do not have a counter part on the other layer to be directly above or below them are referred to as `non-dimer' sites.~In the tight-binding description of BLG, four atomic sites per unit cell with both intra- and interlayer hopping amplitude result in four bands; two conduction bands and two valence bands.~The low energy bands near $K$ and $K^\prime$ valleys are parabolic and not linear as in monolayer graphene~(left panel in Fig.~1d).~This significant difference between the bandstructures of graphene and Bernal BLG can be understood when a displacement field is applied.~While the Dirac-like bands in graphene remain unchanged in the presence of a displacement field, the parabolic band touchings in Bernal BLG are replaced by a finite bandgap with the formation of flatter bands at low energy~\cite{zhang2009direct, jayaraman2021evidence}~(right panel in Fig.~1d).~While the resistance at the charge neutrality point (CNP) in gapped BLG is mainly found to be insulating~\cite{zhang2009direct, jayaraman2021evidence, oostinga2008gate}, evidence of transport via one-dimensional edge modes has also been reported~\cite{PhysRevLett.121.136806}.~Diverging single-particle DOS near the band edge opens up new possibilities for exploring correlated phases in Bernal BLG subjected to a displacement field~\cite{zhang2009direct, oostinga2008gate}, which we shall address later in this review.\\
\begin{figure*}[bth]
\includegraphics[width=1.0\textwidth]{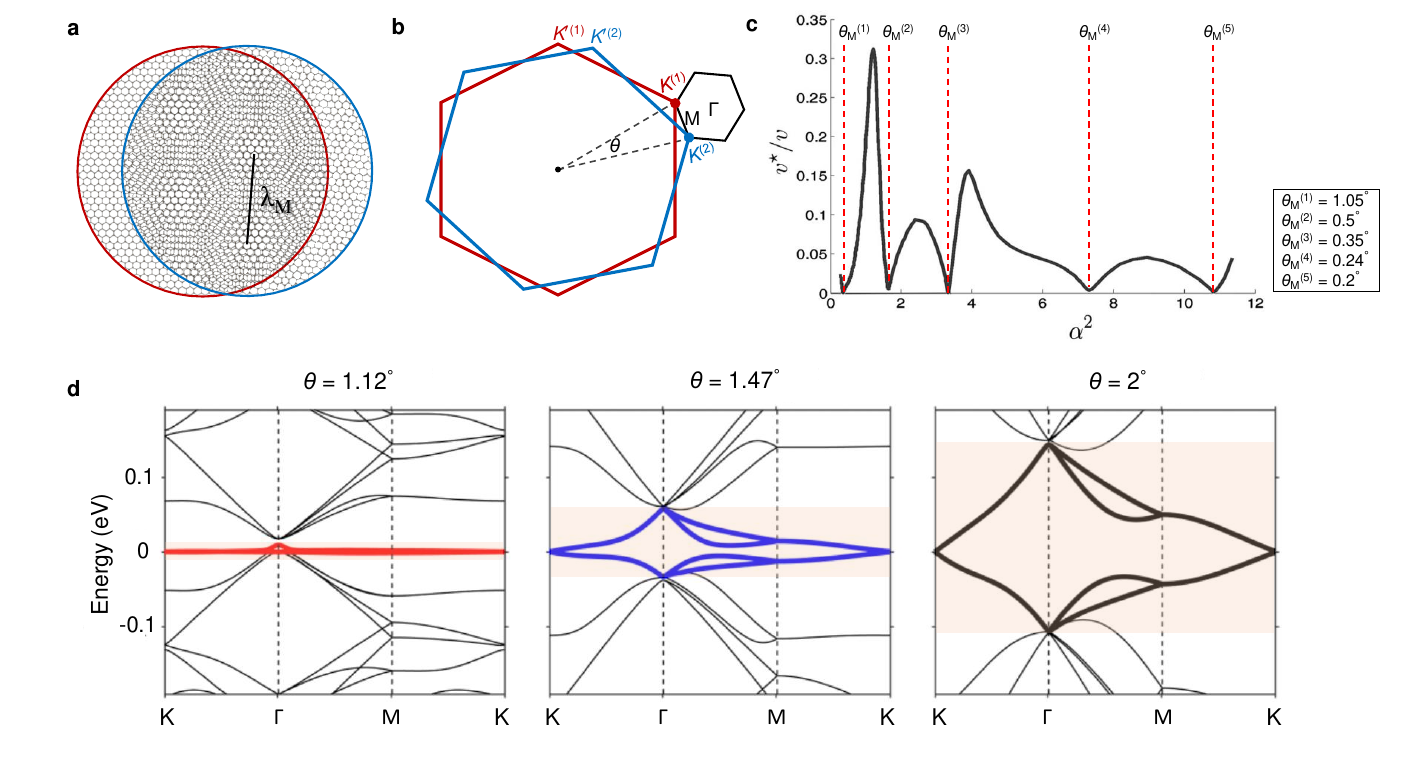}
\captionsetup{justification=raggedright,singlelinecheck=false}
\justify{\textbf{Fig.~2.~Twisted bilayer graphene: Moir\'{e} pattern and flat bands}.~\textbf{a.}~An illustration of moir\'{e} pattern formed when two graphene layers are mutually rotated by a twist angle $\theta$.~The period of the moir\'{e} pattern is $\lambda_{SL}=2/a\sin{(\theta/2)}$, where $a=0.246$~nm is the lattice constant of graphene.~\textbf{b.}~For a small $\theta~(<2\degree$), $\lambda_{SL}=a/\theta>>a$, with a smaller Brillouin zone in the reciprocal space.~This mini Brillouin zone is constructed from the difference between $K$($K^\prime$) vectors of the top (1) and bottom (2) graphene layers.~$\Gamma$ and $M$ are other points in the Brillouin zone.~\textbf{c.}~Renormalised Fermi velocity is plotted as a function of $\alpha^2$, where $\alpha=w/v_Fk_\theta$, $w$ is the interlayer coupling, $k_\theta$ is the momentum displacement between two Dirac cones, and $v_F$ is the Fermi velocity of graphene.~The velocity drastically reduces to zero for $\theta\approx1.05\degree, 0.5\degree, 0.35\degree, 0.24\degree$, $0.2\degree$ which are called the `magic' angles.~The plot is taken with permission from ref.~\cite{bistritzer2011moire}.~\textbf{d.}~The TBG band structure changes significantly with $\theta$.~The bandwidth (shaded by orange colour) of the lowest energy bands decreases when $\theta$ is decreased and becomes extremely flat close to the magic angle. The band structures are taken from ref.~\cite{PhysRevResearch.1.013001}}.
\end{figure*}

\section{Moir\'{e} superlattices with graphene}
\subsection{\normalsize Graphene/Boron Nitride superlattices}

The realisation of correlation-driven phenomena in graphene has remained a longstanding goal, primarily due to the absence of strong electron-electron interactions within its low-energy bands.~It was theoretically proposed that a moir\'{e} pattern with graphene leads to hybridised narrow Bloch bands as a consequence of a large superlattice unit cell~\cite{PhysRevB.90.155406, PhysRevLett.99.256802, bistritzer2011moire, PhysRevResearch.1.013001}.~Moir\'{e} supercells can be artificially created in solid-state systems by either vertically arranging two lattices with a marginal lattice mismatch or by rotating two identical lattices with respect to one another. The first generation of 2D moir\'{e} superlattices was created by aligning hBN to monolayer graphene~\cite{xue2011scanning, yankowitz2012emergence, ponomarenko2013cloning, doi:10.1126/science.1237240}, where a small lattice mismatch between the layers generated a moir\'{e} pattern (Fig.~1e).~The low-energy bands split into multiple electron and hole minibands with finite energy gap between them due to the folded Brillouin zone of the superlattice~\cite{PhysRevB.90.155406}.~Interestingly, a finite gap opening between the first and second minibands leads to insulating states in transport measurements~\cite{yankowitz2012emergence}~(Fig.~1f).~Typically, the observation of a larger band gap on the hole side, compared to the electron side, indicates a significant electron-hole asymmetry within the system.
In a similar manner, when a Bernal BLG is aligned with hBN, a moir\'{e} pattern is formed (Fig.~1g) followed by the opening of a band gap upon completely filling the low-energy bands~\cite{PhysRevB.90.155406}, which are flatter than those in graphene (Fig.~1h).\\

When an electron under the moir\'{e} potential is subjected to a large out-of-plane magnetic field a completely different fractal energy spectrum known as `Hofstadter's butterfly' emerges~\cite{PhysRevB.14.2239}.~The general solution for the motion of electrons leads to a condition of $\phi/\phi_0 = p/q$, where $\phi$ is the magnetic flux per unit cell, $\phi_0=h/e$ is the magnetic flux quantum, $p$ and $q$ are co-prime integers indicating the splitting of a single-particle Bloch band into $q$ subbands.~This fractal energy spectrum is mainly governed by two length scales:~(a)~lattice constant and~(b)~magnetic length that depends on the periodic potential and magnetic field.~Initial attempts at experimentally realising the Hofstadter's butterfly had failed due to the difficulties in reconciling these two length scales.~For instance, in a usual crystal lattice with an interatomic spacing of less than one nanometer, the necessary magnetic field to satisfy Hofstadter's condition is around 10,000~T, which is unfeasible experimentally.~On the other hand, in an artificially engineered superlattice with a period of $\sim 100$ nm~\cite{PhysRevB.43.5192, PhysRevLett.83.2234}, the necessary field becomes too small to overcome the effect of disorder.~The moir\'{e} superlattices formed by stacking a layer of hBN on Bernal bilayer graphene with near $0\degree$ twist angle have a period of $\sim15$~nm, making it an ideal platform to explore the Hofstadter's butterfly spectra in a range of accessible magnetic fields~\cite{dean2013hofstadter}. Despite the observation of additional band-insulating states and Hofstadter's butterfly spectra driven by the superlattice unit cell, the exploration of strongly correlated physics is limited in hBN/graphene moir\'{e} systems due to the absence of flat bands and strong electron-electron interactions. As a result, we will not delve into these systems extensively. However, it is worth noting that the hBN/graphene moir\'{e} superlattice potential can significantly impact the magnetic phases discussed later in this article. Interestingly, contrary to graphene/hBN superlattice, the presence of a very flat cubic band at the $K$-points of the Brillouin zone in ABC-trilayer graphene (ABC-TLG) leads to magnified electronic correlations~\cite{chen2019evidence, chen2020tunable}, particularly when aligned with hBN. The bandwidth can be significantly reduced using a displacement field, resulting in a diverging DOS.

\begin{figure*}[bth]
\includegraphics[width=1.0\textwidth]{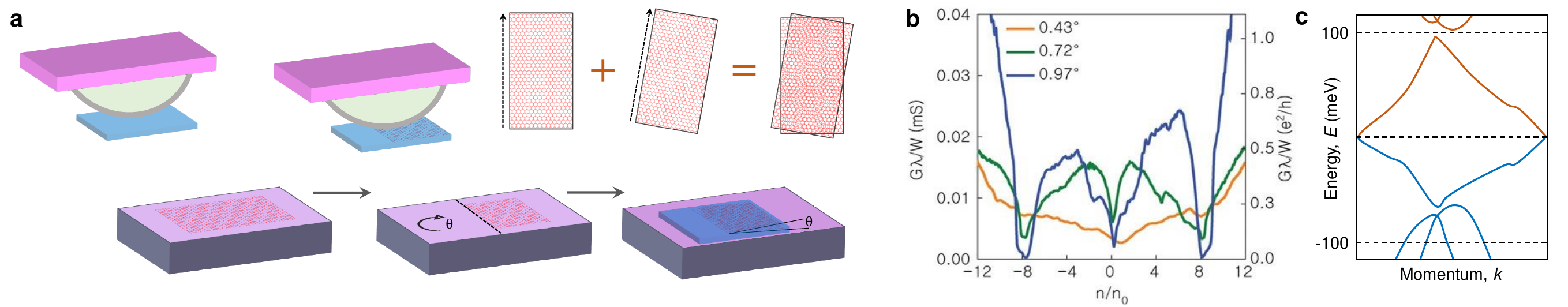}
\captionsetup{justification=raggedright,singlelinecheck=false}
\justify{\textbf{Fig.~3.~Twisted bilayer graphene: Fabrication and transport measurements}.~\textbf{a.}~Schematic of the `tear and stack' technique for assembling TBG heterostructures. Polydimethylsiloxane (PDMS) hemisphere (light green colour) coated with an adhesive polymer (gray colour), mounted on a transparent glass slide (pink), is used to pick the 2D layers from the substrate. First, a hBN flake with a sharp edge is picked up. This sharp edge is then used to tear a graphene flake, followed by picking up the first half and leaving the other half on the substrate. Next, the sample stage is rotated by a desired angle, and the second half is picked up. Finally, the heterostructure is released on a substrate.~\textbf{b.}~Normalised conductance $G\lambda/W$ plotted as a function of filling $n/n_0$ for three different TBG samples where $\lambda$ is moir\'{e} period, $W$ is the width of the sample, $n$ is the carrier density, and $n_0$ is the carrier density corresponding to one carrier per unit cell. Data adapted from ref.~\cite{kim2017tunable}. Note that this initial work assumed a total of eight carriers in a fully filled unit cell. However, subsequent experiments considered it to be four, assigned to two spins and two valleys. The conductance showed two prominent minima on either side of the CNP. The twist angle $\theta$ in the TBG samples is typically calculated using the relation, $n_s=8\theta^2/\sqrt{3}a^2$ where $a = 0.246$ nm is the lattice constant of graphene and $n_s=4n_0$ is the charge carrier density corresponding to a fully filled superlattice unit cell, implying the position of identical conductance minima on the electron and hole sides. For transport data in the magic angle regime, refer to Fig.~6.~\textbf{c.}~Recreated schematic of the band structure obtained using the tight-binding model for TBG with $\theta=1.8\degree$ in ref.~\cite{PhysRevLett.117.116804}. The conduction and valence bands are denoted by orange and blue colours. The low-energy bands become less dispersive compared to monolayer graphene, accompanied by the formation of finite gaps between the first and second bands. Note that the bandwidth is reduced to $\approx 150$ meV.}
\end{figure*}

\subsection{\normalsize Twisted Bilayer Graphene}

In this section, we focus on the formation of superlattices by stacking two graphene layers with a small twist angle between them (Fig.~2a-b). Theoretical studies of twisted bilayers of graphene date back to 2007, when it was first proposed using the continuum model that the Fermi velocity is substantially reduced in the small twist angle regime~\cite{PhysRevLett.99.256802}. Such a strong reduction indicates the formation of narrower bands, where electron-electron interactions are expected to be significantly amplified due to their close proximity within a small energy window.~Exotic phases can be stabilised in such a system, particularly when the Fermi energy encounters a saddle point in the energy dispersion, known as a van Hove singularity (vHS), where the DOS diverges~\cite{PhysRev.89.1189}. In monolayer graphene, the position of the vHS at $\sim2.8$ eV is quite high compared to the experimental  Fermi energy scales, making it inaccessible. However, the flat bands in TBG are of the order of a few tens of meVs, whereby the vHSs can be accessed via appropriate electrostatic gating. In one of the earliest experiments, scanning tunneling spectroscopy on graphite samples with a top graphene layer rotated by a small angle ($1.79\degree$) demonstrated two prominent peaks in the local DOS, suggesting the presence of vHSs in the conduction and valence bands~\cite{li2010observation}.~In the following years, theorists made further progress by reconstructing the continuum model Hamiltonian considering different strengths of interlayer hopping to obtain the band structure~\cite{PhysRevB.76.085425, PhysRevB.78.045405, PhysRevB.77.155416, doi:10.1021/nl902948m, PhysRevLett.101.056803, PhysRevB.81.161405, PhysRevB.81.245412, PhysRevB.82.121407}. In their seminal paper in 2011~\cite{bistritzer2011moire}, Rafi Bistritzer and Allan H. MacDonald predicted that when two graphene layers are rotated by specific angles called `magic' angles, the low-energy bands become extremely flat, and the Fermi velocity of the charge carriers reduces to zero (Fig.~2c). The first magic angle was predicted to be at 1.05$^\circ$. Their proposal of perfect flat bands stimulated numerous experimental studies in the following years, including scanning tunneling spectroscopy~\cite{PhysRevLett.106.126802, PhysRevLett.109.126801, PhysRevLett.121.037702} and magnetotransport measurements on graphene layers mutually rotated by twist angles slightly larger than the magic angle~\cite{PhysRevLett.108.076601, PhysRevLett.110.096602, mahapatra2017seebeck, PhysRevLett.125.226802}. While the linear dispersion is preserved and hence the interlayer coupling remains weak, there exists twice the number of Dirac cones due to the two layers. Although the layers can be tuned independently via electric and magnetic fields, the system cannot be modeled as two monolayers in parallel, and therefore, the layer interactions cannot be neglected~\cite{PhysRevLett.108.076601}. Apart from in-plane transport measurements, interlayer electrical resistance measurements demonstrated a strongly insulating behaviour at CNP, with significantly higher values compared to Bernal BLG~\cite{PhysRevLett.110.096602, mahapatra2017seebeck}. To explain this phenomenon, theoretical calculations suggested that a significant rotational mismatch between the graphene layers promotes incoherent tunneling, which in turn accounts for the observed giant interlayer resistance~\cite{PhysRevLett.111.066803}.\\

The experimental and theoretical findings mentioned above suggest that precise control over the twist angle between the two graphene layers is crucial for realising correlated physics, as is evident from the evolution of the band structure with twist angle (Fig.~2d). Despite rigorous experimental efforts for several years, the `magic' angle regime was not realised experimentally, primarily because the fabrication methods in many of these early works~\cite{PhysRevLett.108.076601, PhysRevLett.110.096602} relied on the vertical stacking of two separately exfoliated graphene layers.~Twist control between the layers was achieved by rotating the sharp edge of one layer with respect to the other, as exfoliated graphene flakes typically have straight edges that are either zigzag or armchair in configuration~\cite{you2008edge}. 

\begin{figure*}[bth]
\includegraphics[width=1.0\textwidth]{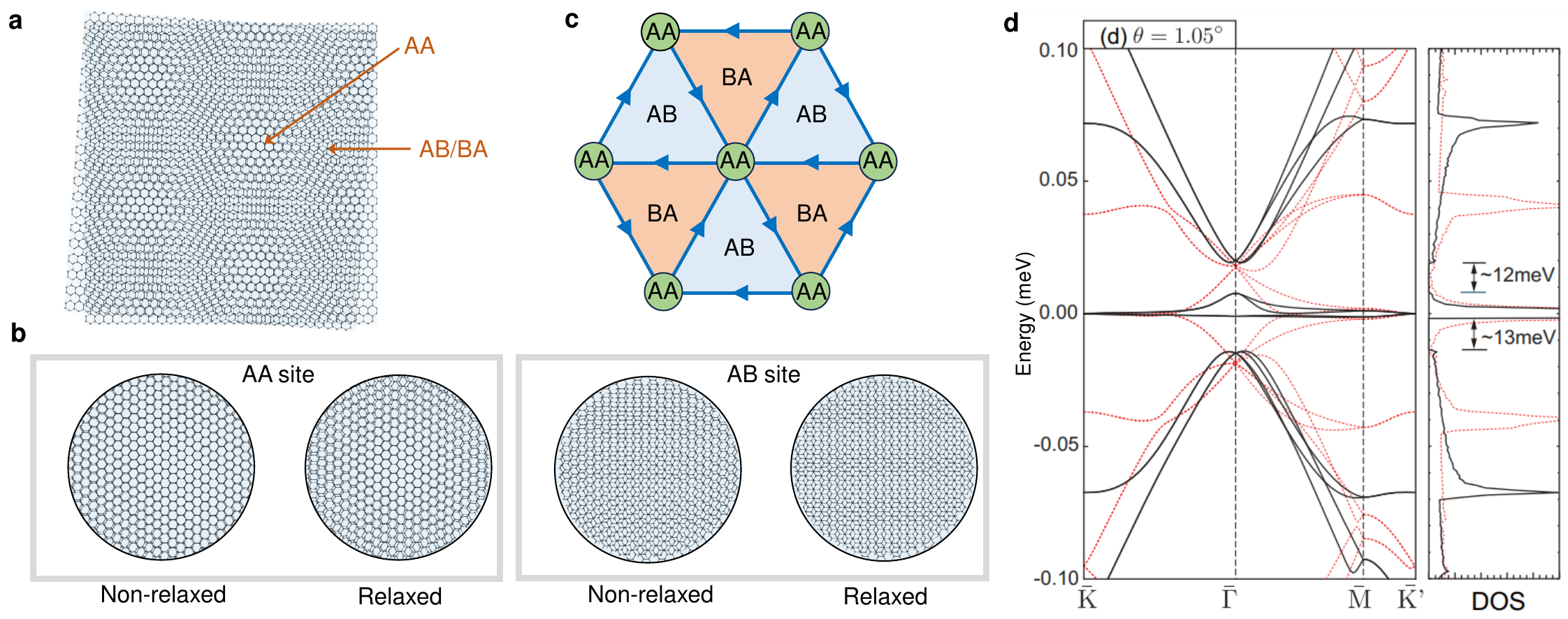}
\captionsetup{justification=raggedright,singlelinecheck=false}
\justify{\textbf{Fig.~4.~Lattice relaxation in twisted bilayer graphene}.~\textbf{a.}~The moir\'{e} pattern formed when the twist angle is $4\degree$.~The arrows represent different regions with stacking order AA, AB, and BA.~\textbf{b.}~The effect of lattice relaxation in the system leads to a change in the area of AA and AB/BA regions.~In the relaxed structure, the area of AA region minimises and the area of AB/BA region maximises.~\textbf{c.}~Schematic of the triangular network of AB (light blue)/BA (orange) domain walls in marginally TBG. The blue arrows along the domain walls indicate the direction of 1D edge modes. AA domains are represented by small green circles. AB/BA domains can be gapped out by applying an out-of-plane electric field, and transport occurs via the edge modes.~\textbf{d.}~Band structure and DOS of TBG in relaxed (black solid lines) and nonrelaxed (red dashed lines) configurations.~For the non-relaxed structure there is no bandgap between the first and second band; however, a finite gap opens up in case of the relaxed structure.~In addition, the width of the flat band in the relaxed configuration is lower than the non-relaxed configuration.~Band structure taken with permission from ref.~\cite{PhysRevB.96.075311}.}
\end{figure*}

However, this method has limitations due to challenges in precisely identifying the crystallographic axes of the graphene layers from their edges, as the edges often exhibit a mixture of zigzag and armchair configurations.~Consequently, achieving the desired twist angle reliably turned out to be a challenging task, with the best devices being at $\sim 2^\circ$ twist angle.~In 2016, as a remarkable improvement over existing methods, E. Tutuc and colleagues proposed a novel technique, now popularly called the `tear and stack', where both the top and bottom graphene layers were obtained from the same graphene flake~\cite{doi:10.1021/acs.nanolett.5b05263}.~The strong van der Waals interactions between hBN and graphene enabled the tearing of the graphene flake along the sharp edge of hBN, followed by picking up and rotating the first and second halves successively to form the TBG (Fig.~3a). The identical crystallographic orientation of the two graphene layers provides much better control over the twist angle between them. Soon after, in two independent works~\cite{PhysRevLett.117.116804, kim2017tunable}, electrical transport measurements in low angle TBG ($\sim 0.4^\circ-2^\circ$) revealed two additional insulating peaks symmetrically located around the CNP on the electron-doped and hole-doped sides (Fig.~3b).~These newly observed insulating states were attributed to the opening of band gaps between the first and second moir\'{e} bands, very similar to observations in graphene/hBN moir\'{e} lattices~\cite{yankowitz2012emergence, dean2013hofstadter, ponomarenko2013cloning, doi:10.1126/science.1237240}. The calculated band structure for the enlarged superlattice unit cell (folded Brillouin zone) clearly exhibited a band gap consistent with the experimental results (Fig.~3c). Subsequently, in 2018, P. Jarillo-Herrero's group at MIT successfully achieved the magic-angle condition by setting a twist angle of $1.05\degree-1.16\degree$ between the two graphene layers. They discovered strong insulating states at the half-filling of the bands~\cite{cao2018correlated}. Even more surprisingly, when slightly doped away from half-filling, the resistance abruptly dropped to zero upon cooling the sample below a temperature of $\sim 1$ Kelvin, a hallmark of superconductivity~\cite{cao2018unconventional}. While the insulating states were expected due to strong electron-electron interactions within the half-filled flat bands, the observation of superconductivity was entirely unexpected. These observations provided a breakthrough and sparked a significant number of theoretical and experimental investigations into the correlated phenomena in MATBG. We will discuss these experimental signatures in detail in later sections. In the remainder of this section, we will focus on the band structure of TBG near the magic angle, as it will be referred to multiple times in this review.\\

The flattening of the bands as the twist angle decreases can be qualitatively understood as a consequence of the competition between the intra- and interlayer coupling strength.~At the magic angles, the interlayer hybridisation energy becomes comparable to the energy difference between the Dirac cones of two graphene layers~\cite{PhysRevLett.99.256802, bistritzer2011moire} and the lowest of the hybridised states is pushed towards zero energy~(Fig.~2d).~The flat bands in momentum space imply localisation of electrons in real space.~The moir\'{e} lattice can be visualised as a combination of AA and AB/BA regions (Fig.~4a). The calculations of the density profile show that electrons are highly concentrated in AA stacking regions, whereas AB/BA regions have a lower electron density~\cite{PhysRevB.84.195421, li2010observation}.
\begin{figure*}[bth]
\includegraphics[width=1.0\textwidth]{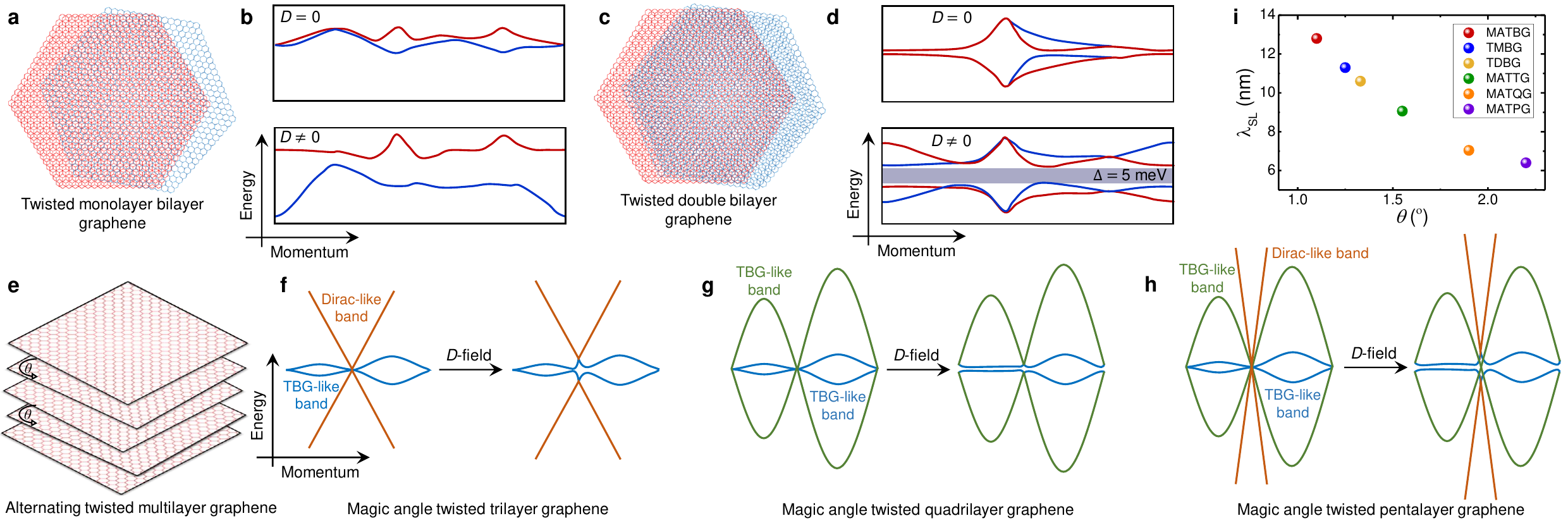}
\captionsetup{justification=raggedright,singlelinecheck=false}
\justify{\textbf{Fig.~5.~The family of twisted graphene systems}.~The resultant moir\'{e} patterns and corresponding schematic of the band structures for \textbf{a,b} twisted monolayer bilayer graphene (TMBG), \textbf{c,d} twisted double bilayer graphene (TDBG), and \textbf{e-h} alternating twisted multilayer graphene systems such as magic angle twisted trilayer, quadrilayer and pentalayer graphene (MATTG, MATQG, and MATPG), respectively. In all these cases, the application of a displacement field flattens the bands and opens a band gap. The scale of energy for the recreated schematic of the band structure of TMBG and TDBG are (-40, 40) meV and (-10, 10) meV, presented in ref.~\cite{polshyn2020electrical} and~\cite{cao2020tunable}, respectively.~An infinite class of moir\'{e} systems can be formed by stacking graphene layers with alternating twist angles between the adjacent layers .~To date, a maximum of five graphene layers have been combined together to fabricate twisted multilayer systems. Alternating twisted multilayers feature TBG-like bands denoted in blue/green colour that are relatively flat, along with a linear Dirac-like band denoted by orange colour (for odd $m$ such as MATTG and MATPG).~The original band structure is presented in ref.~\cite{park2022robust, zhang2022promotion} ~\textbf{d.}~The variation of superlattice period $\lambda_{SL}$ with the twist angle $\theta$ where flat bands are expected for different graphene moir\'{e} systems.}
\end{figure*}
The band structure calculations in TBG rely on the assumption that the honeycomb lattice structure of graphene is rigid and it does not change when two graphene layers are stacked with a small rotation~\cite{PhysRevLett.99.256802, bistritzer2011moire, PhysRevB.85.195458}.~However, this is not the case in a real system where the lattice structure relaxes to an energetically favourable configuration~\cite{PhysRevB.84.045404, van2015relaxation, PhysRevB.96.075311}.~The interlayer energy scale associated with different regions (AA, AB/BA) determines the final lattice structure after atomic relaxation.~Since the interlayer binding energy of AA regions is much higher than AB/BA regions, the area of AA regions minimises while the area of AB/BA regions maximises in the relaxed lattice structure~\cite{PhysRevB.96.075311, PhysRevB.98.235137, yoo2019atomic}, modifying the original moir\'{e} pattern (Fig.~4b). Such atomic reconstructions favouring interlayer commensurability have been directly imaged using transmission electron microscopy~\cite{yoo2019atomic} and scanning tunneling microscopy~\cite{PhysRevLett.125.236102, PhysRevLett.121.037702}.~We also note that the AB/BA regions can be gapped out using a perpendicular electric field, leading to a triangular network of chiral one-dimensional~(1D)~modes~(Fig.~4c).~Such 1D modes are prominent in marginally twisted samples with $\theta \sim 0.1-0.5\degree$, where interesting phenomena such as Aharanov Bohm effect and Fabry-Perot oscillations have been observed~\cite{xu2019giant, doi:10.1021/acs.nanolett.2c00627, doi:10.1021/acs.nanolett.8b02387}.~The band structure and DOS for non-relaxed (red dashed line) and relaxed (black solid line) configurations presented in Fig.~4d show two primary differences.~First, in case of a relaxed TBG, a finite energy gap opens up between the first and second moir\'{e} bands while it almost vanishes for the non-relaxed structure.~Second, the lowest energy bandwidth decreases with twist angle in both cases.~However, the flattening of the bands is slower in the relaxed structure.~Therefore, at the same angle, the bandwidth in the relaxed configuration is larger than its non-relaxed counterpart, lowering the critical angle where the Fermi velocity reduces to zero in the relaxed TBG.~These two results are in good agreement with the experiments discussed later in this article.

\subsection{\normalsize Twisted Multilayer Graphene}

The enhanced correlations within the low-energy flat bands in MATBG have stimulated extensive theoretical and experimental progress in search of other systems that exhibit similar properties but differ in certain details. Identifying such systems offers several advantages, including expanding the family of moir\'{e} materials, facilitating easier fabrication, and potentially providing a tunable phase diagram. In this section, we introduce the general class of multilayer moir\'{e} graphene systems that share similar phases with TBG but vary in terms of their band structures, symmetries, and interaction strengths.\\

When a Bernal BLG is slightly rotated with respect to monolayer graphene, a superlattice called twisted monolayer bilayer graphene (TMBG) can be formed~\cite{polshyn2020electrical, PhysRevB.102.035411, xu2021tunable, chen2021electrically, li2022imaging, PhysRevResearch.2.033150, he2021competing, PhysRevLett.128.126401, polshyn2022topological}~(Fig.~5a).~The breaking of two-fold rotational symmetry $C_2$ and mirror symmetry $M_y$ results in a lower crystal symmetry in TMBG. The parabolic bands in Bernal BLG strongly hybridise with the Dirac-like bands in monolayer graphene when a displacement field is applied (Fig.~5b). A similar system composed of two rotated sheets of Bernal BLG is known as twisted double bilayer graphene (TDBG)~\cite{cao2020tunable, liu2020tunable, shen2020correlated, he2021symmetry, PhysRevLett.123.197702, PhysRevB.99.235417, PhysRevB.99.235406, doi:10.1021/acs.nanolett.9b05117, lee2019theory, rickhaus2021correlated, PhysRevLett.128.057702, doi:10.1021/acs.nanolett.1c03066, PhysRevLett.125.176801, PhysRevB.101.125428, wang2022bulk, PhysRevLett.126.026801, sinha2020bulk} (Fig.~5c). When subjected to a finite displacement field, the bands become narrower, leading to enhanced Coulomb interactions (Fig.~5d).
\begin{figure*}[bth]
\includegraphics[width=1.0\textwidth]{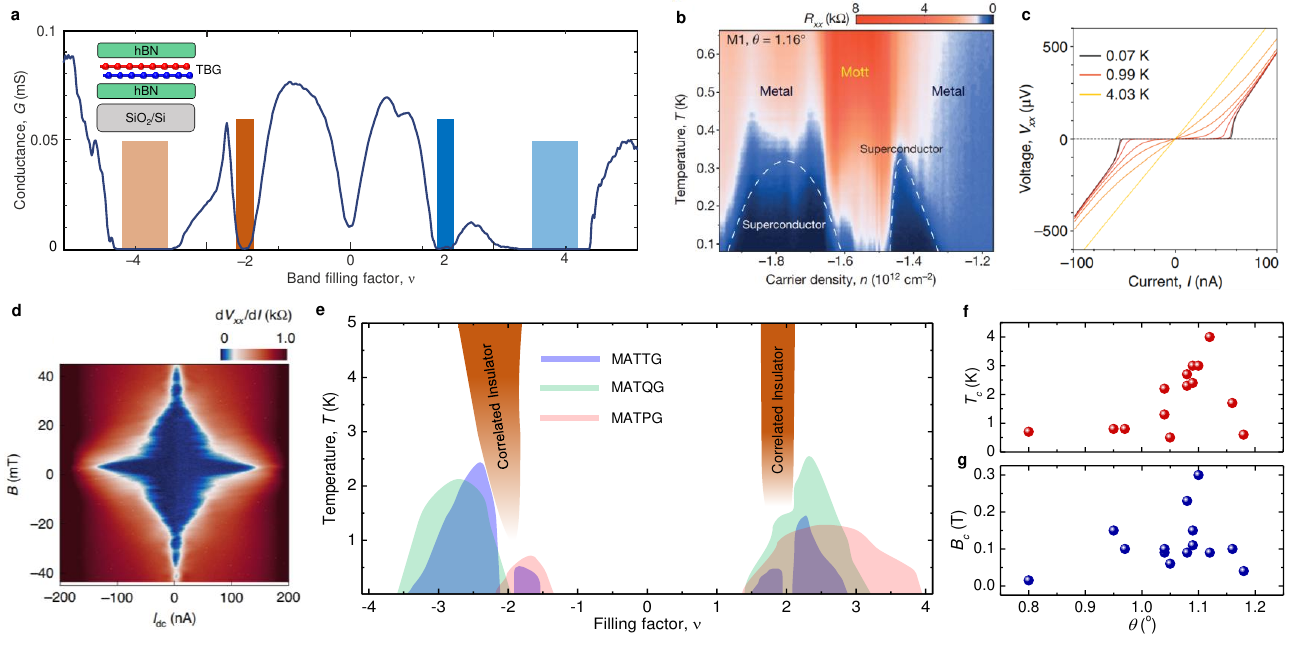}
\captionsetup{justification=raggedright,singlelinecheck=false}
\justify{\textbf{Fig.~6.~Correlated insulators and superconductivity in various twisted graphene systems}.~\textbf{a.}~Electrical conductance $G$ measured by sweeping the density $n$ across the flat bands for a device with $\theta=1.08\degree$ at a temperature of $T=0.3$~K.~Data taken with permission from ref.~\cite{cao2018correlated}~The minima in $G$ appear at band filling factor of $\nu=0$, $\nu=\pm4$, which are at the full fillings (lighter-shaded region) of the conduction $(+)$ and valence $(-)$ bands, and at $\nu=\pm2$ which are at the half-fillings (darker-shaded region) of the conduction $(+)$ and valence $(-)$ bands.~\textbf{b.}~Four probe longitudinal resistance $R_{xx}$ plotted as a function of carrier density near the half-filling of the valence band, and temperature.~Two superconducting domes outlined by white dashed arrows are seen on either side of the correlated insulator that appears at $n=-1.5\times10^{12}$~cm$^{-2}$, corresponding to $\nu=-2$.~The highest critical temperature $T_c$ of the superconductor reported by the authors is about $T=0.5$~K (at $50\%$ of the normal-state resistance).~Data taken with permission from ref.~\cite{cao2018unconventional}.~\textbf{c.}~Current-voltage ($I-V_{xx}$) characteristics measured at a density where superconductivity is observed.~At the lowest $T=0.07$~K, $I-V_{xx}$ curve is perfectly non-linear with a critical current of about 50~nA.~As the temperature increases the superconductivity becomes weaker leading to a linear curve at $T=4.03$~K where superconductivity completely disappears.~Data taken with permission from ref.~\cite{cao2018unconventional}.~\textbf{d.}~Differential resistance d$V_{xx}$/d$I$ plotted as a function of dc bias current $I_{dc}$ for different perpendicular $B$ for $\theta=1.12\degree$.~Data taken with permission from ref.~\cite{saito2020independent}.~\textbf{e.}~A summary of the correlated insulators and superconductivity observed in alternating twisted multilayer graphene systems~\cite{park2021tunable, park2022robust, zhang2022promotion}.~Blue, green, and red colours denote superconducting domes in MATTG, MATQG, and MATPG, respectively.~Among these systems, MATTG shows correlated insulators near $\nu=\pm2$ at a higher temperature where superconductivity disappears; correlated insulators are obscured by the superconducting domes at lower temperatures.~MATQG, and MATPG do not exhibit correlated insulating states; however, the density range over which superconductivity appears is surprisingly larger with increasing the number of layers in the system.~\textbf{f.-g.}~Superconducting $T_c$ and out-of-plane critical field $B_c$ obtained from ref.~\cite{cao2018unconventional, lu2019superconductors, cao2021nematicity, saito2020independent, stepanov2020untying, arora2020superconductivity, liu2021tuning}. $T_c$ was calculated from the transition to $50\%$ of the normal state resistance. Although $T_c$ and $B_c$ both can vary widely over samples, they are seen to be maximum near $\theta\approx1.1\degree$.}
\end{figure*}
Additionally, in the presence of a displacement field, the degeneracy of the low-energy bands in both TMBG and TDBG is lifted due to the finite gap opening at the CNP.\\

Apart from strong correlations in twisted graphene heterobilayer superlattices, an infinite class of multilayer graphene systems has been theoretically predicted to harbour flat bands at specific magic angles~\cite{PhysRevB.100.085109, PhysRevX.9.031021, PhysRevLett.128.176403, tritsaris2020electronic, doi:10.1021/acs.nanolett.3c00275, park2022robust, burg2022emergence, zhang2022promotion}. These systems are known as alternating twisted multilayer graphene since the relative twists between two neighbouring layers have the same magnitude, but alternate in sign (Fig.~5e). In a system with $m$ graphene layers, several moir\' {e} patterns emerge with $m/2$ TBG-like flat bands if $m$ is even, and $(m-1)/2$ TBG-like bands, along with 1 Dirac-like band if $m$ is odd~\cite{PhysRevB.100.085109, PhysRevX.9.031021}. For odd $m$, the system possesses an out-of-plane mirror symmetry $M_z$, which is substituted by a two-fold in-plane rotational axis $C_2$ for even $m$. The band structure near zero energy for all these systems is strongly hybridised under a displacement field (Fig.~5f-h). Twisted trilayer graphene has been of particular interest due to its intriguing and rather robust superconducting phase~\cite{park2021tunable, hao2021electric, kim2022evidence, liu2022isospin, shen2023dirac, PhysRevLett.127.166802, lin2022zero, cao2021pauli, turkel2022orderly, PhysRevLett.127.097001, PhysRevB.104.L121116, pantaleon2023superconductivity, fischer2022unconventional, PhysRevX.12.021018}. Fig.~5i illustrates how the moir\' {e} wavelength varies for different multilayer systems at different twist angles where flat bands are expected. The common thread linking all these superlattices is the presence of flat bands near the Dirac points. Various lattice and band symmetries determine experimentally tunable parameters that control electron-electron interactions. Although rapid progress in experiments has facilitated the creation of more complex heterostructures, no other system currently exhibits as many correlated phases as MATBG.

\section{Emergent Properties}

\subsection{\normalsize Correlated Insulators and Superconductivity}

The discovery of correlated insulators~\cite{cao2018correlated} and superconductivity~\cite{cao2018unconventional} elicited extensive interest in exploring a rich phase diagram in MATBG. In this section, we focus on the experimental signatures of insulating and superconducting states in MATBG, which have now been investigated by several research groups~\cite{yankowitz2019tuning, lu2019superconductors, polshyn2019large, choi2019electronic, jaoui2022quantum, codecido2019correlated, saito2020independent, stepanov2020untying}.~Before describing the observed physics, it is essential to define the term `band filling' as it plays a crucial role in identifying the correlated states within the flat bands.~In TBG, the combination of spin ($\uparrow, \downarrow$) and valley ($K$, $K^\prime$) degrees of freedom results in a total degeneracy of $4$.~Hence, the carrier density corresponding to the complete filling of the first conduction ($+$)/valence ($-$) band is the superlattice density $n_s=\pm4/A$, where $A$ is the area of the moir\'{e} unit cell. The twist angle $\theta$ in TBG samples is typically calculated using the relation, $n_s=8\theta^2/\sqrt{3}a^2$ where $a = 0.246$ nm.~The band filling is expressed as $\nu = n/n_0$, where $n_0$ is the density of one electron/hole in the unit cell.~Therefore, $\nu=\pm1, \pm2, \pm3$ and $\pm4$ indicate quarter, half, three-quarter, and full-filling of the conduction ($+$)/valence ($-$) band, respectively.~The first observation of insulating states at $\nu=\pm2$ is markedly non-trivial since a half-filled band is expected to show metallic behaviour~(Fig.~6a).~To understand the origin of the insulating states, we briefly discuss the Hubbard model~\cite{hubbard1964electron}, which is used as a powerful tool to solve the Hamiltonian of correlated electrons in a solid.~This model was initially applied to explain the itinerant magnetism in transition metals like iron and nickel~\cite{PhysRevA.82.061603, PhysRevB.56.3159}.~The Hubbard Hamiltonian is constructed using two energy scales associated with the motion of electrons in a solid.~First, the kinetic energy of electrons which essentially governs the hopping of electrons from one site to another is determined by the hopping energy $t$.~Secondly, the strength of the Coulomb interaction between electrons is defined by $U$.~In the limit of $U/t>>1$, hopping of electrons between two neighbouring sites is forbidden due to a large Coulomb repulsion between electrons that disfavours two electrons on the same site. At half-filling of the band, all lattice sites are therefore singly occupied, leading to a correlated insulator.~In the case of flat bands in realistic MATBG samples~\cite{cao2018correlated, cao2018unconventional, lu2019superconductors, yankowitz2019tuning}, the Coulomb repulsion $U\approx120$~meV and the kinetic energy of the electrons $t\approx90$~meV, with the system being in the moderate limit of the Hubbard model, $U/t\sim 1$.~While we note that these states at $\nu=\pm2$ are usually insulating with a thermally activated gap of $\sim0.3$~meV~\cite{cao2018correlated, lu2019superconductors, yankowitz2019tuning}, semi-metallic behaviour have also been observed~\cite{arora2020superconductivity, bhowmik2022broken}~Remarkably, on lowering the temperature below 1 Kelvin and slightly doping these insulators away from $\nu=\pm2$, an unexpected phenomenon occurs: the resistance abruptly drops (Fig.~6b), signaling the onset of a superconducting phase~\cite{cao2018unconventional, lu2019superconductors, yankowitz2019tuning, saito2020independent, stepanov2020untying}.\\

We now highlight some of the experimental signatures based on which superconductivity has been claimed in various twisted graphene systems~\cite{cao2018unconventional, lu2019superconductors, yankowitz2019tuning, park2021tunable, cao2021pauli, zhang2022promotion, park2022robust}.~This is important in order to understand the evolution of the field and how the quality of the experimental data has substantially improved with time.~Notably, the `zero' resistance of superconductors has not been observed in some of these studies, limited by contact resistances~\cite{codecido2019correlated, liu2020tunable, wu2021chern}.~Alongside the drop in resistance, the presence of sharply-switching non-linear current-voltage $I-V$ characteristics have been interpreted as evidence for superconductivity (Fig.~6c).~Furthermore, measurements in the presence of an out-of-plane magnetic field were employed to investigate the Meissner effect, whereby a superconductor screens an external magnetic field.~The characteristic length scale governing the Meissner effect is the London penetration depth $\lambda$ that determines the decay of the externally applied magnetic field inside the superconductor.~However, 2D superconductors are fundamentally different from their three-dimensional counterparts.~In case of an ultrathin superconductor where the thickness of the film $d$ is smaller than $\lambda$, the spatial distribution of the magnetic field inside the superconductor is determined by the Pearl length $\Lambda=2\lambda^2/d>>\lambda$~\cite{pearl1964current}.~In MATBG, the thickness of $\approx1$~nm results in Pearl length that can become much larger than the dimension of the device itself.~Therefore, the screening current which generates the self-field inside the superconductor cannot expel the external magnetic field and a finite field penetration through the bulk occurs.~Therefore, the primary difficulty in identifying superconductivity in MATBG arises because of the absence of a detectable Meissner effect.~In moir\'{e} systems, the measurement of differential resistance as a function of critical current and magnetic field  shows periodic oscillations, that have been interpreted as  Fraunhofer oscillations arising from unintentional Josephson junction weak links in the sample. These oscillations signifying quantum phase coherent transport provide one of the strongest evidences for superconductivity~\cite{cao2018unconventional, lu2019superconductors, yankowitz2019tuning, saito2020independent, stepanov2020untying}~(Fig.~6d). To date, a handful of reports have claimed superconductivity in MATBG with twist angles within a narrow range of $0.98\degree-1.18\degree$\cite{cao2018unconventional, lu2019superconductors, yankowitz2019tuning, saito2020independent, stepanov2020untying, cao2021nematicity}.~Most strikingly, the superconductor emerges as two domes flanking the insulating state, a signature that has been the hallmark of unconventional superconductors such as the high-$T_c$ cuprates~\cite{norman2003electronic}.\\

Superconductivity in MATBG is very unique in comparison to the other known superconductors because of its unprecedented tunability with several experimental parameters.~For instance, the application of hydrostatic pressure can tune the interlayer coupling between two graphene layers and lead to superconductivity at a twist angle away from the magic angle range~\cite{yankowitz2019tuning}.~Furthermore, the dielectric environment was found to play a crucial role in stabilising many correlated phases in moir\'{e} graphene systems.~Recent experiments have demonstrated superconductivity below the magic angle range ($0.98\degree-1.18\degree$) in TBG proximitised by a monolayer tungsten diselenide (WSe$_2$)~\cite{arora2020superconductivity}.~Graphene is known to have a negligible spin-orbit coupling (SOC); the addition of WSe$_2$, therefore, introduces a finite proximity-induced SOC in TBG.~However, the mechanism of superconducting phase transition in non-magic angle TBG coupled to WSe$_2$ remains enigmatic. As will be discussed later in this review, TBG also exhibits magnetism, believed to be orbital in nature. Surprisingly, the magnetic samples do not exhibit superconductivity even at the established range of magic angles~\cite{serlin2020intrinsic, lin2022spin}, and this remains another open puzzle.\\

In TBG devices, the application of an unbalanced electric potential using top and bottom gates leads to independent control of both carrier density and out-of-plane electric field $D$.~The characteristic parameters of MATBG superconductor, such as the transition temperature and the critical magnetic field have not shown any notable dependence on $D$.~However, the effect of $D$-field becomes extremely important in case of alternating twisted multilayer graphene. Soon after the discovery of superconductivity in MATBG, superconductivity in magic-angle twisted trilayer graphene (MATTG) was demonstrated~\cite{park2021tunable, cao2021pauli, hao2021electric}. This material offers enhanced tunability of both the electronic structure and superconducting properties compared to MATBG. At zero $D$-field, MATTG with a twist angle of approximately $\sqrt{2}\theta_M\approx1.6\degree$ possesses a set of flat bands similar to TBG, along with graphene-like Dirac bands that remain unhybridised due to mirror symmetry~\cite{PhysRevB.87.125414, PhysRevLett.125.116404, PhysRevB.104.L121116}.~However, a significant hybridisation between the flat bands and the Dirac cone occurs under a finite $D$ field. This band modulation with $D$-field is evident in the transport data, where correlated insulators predominantly appear at finite $D$-fields~\cite{park2021tunable, cao2021pauli, hao2021electric, liu2022isospin, shen2023dirac}.~While superconductivity persists on both the electron and hole sides at zero $D$-field, additional superconducting domes emerge at high $D$-fields~\cite{park2021tunable, cao2021pauli, hao2021electric}.~Although MATBG and MATTG both exhibit superconductivity, their electromagnetic field responses show substantial differences.~Most importantly, the superconductivity in MATTG survives in a parallel magnetic field which is 2-3 times larger than the Pauli limit of a Bardeen–Cooper–Schrieffer (BCS)-type superconductor, suggestive of a spin-triplet superconducting state~\cite{cao2021pauli}.~In a conventional superconductor, the application of a magnetic field breaks the degeneracy between opposite spins, leading to the disruption of Cooper pairs and a transition from the superconducting to the normal state~\cite{sacepe2011localization, PhysRevB.56.3372}.~However, in the case of a spin-triplet state, the electrons remain unaffected by the magnetic field since the spins within the Cooper pairs align in the same direction~\cite{maeno2011evaluation, PhysRevLett.104.137002}.~In MATTG, a re-entrant superconducting state was found to exist at a very large in-plane magnetic field reminiscent of certain 3D spin-triplet superconductors~\cite{cao2021pauli}.~To comprehend the mutual connection between MATBG and MATTG in terms of superconductivity, studies were performed on magic-angle twisted quadrilayer graphene (MATQG) and magic angle twisted pentalayer graphene (MATPG).~Surprisingly, MATQG and MATPG both demonstrate superconductivity~\cite{park2022robust, zhang2022promotion}.~Notably, as the number of graphene layers increases in the system, the size of the superconducting domes in carrier density expands (Fig.~6e).~This finding suggests an important contribution of band structure to the emergence and stability of superconductivity in these structures.~Although alternating twisted multilayer graphene systems are microscopically different in terms of symmetries, flat band correlations and the number of graphene layers (even or odd), their transport data unveil a broader family of superconducting materials beyond MATBG. It is to be noted in this context that, in a recent study, superconductivity was observed in TDBG proximitised with WSe$_2$~\cite{su2023superconductivity}.~However, to date, superconductivity has not been demonstrated convincingly in regular TDBG~\cite{liu2020tunable} or TMBG, although electric-field tunable correlated insulators~\cite{chen2021electrically, cao2020tunable, he2021symmetry, liu2022isospin}, ferromagnetism~\cite{polshyn2020electrical, kuiri2022spontaneous, he2021competing, polshyn2022topological}, and magnetic field-driven Chern insulators~\cite{polshyn2020electrical, polshyn2022topological, liu2022isospin} have been reported in both systems over a range of twist angles.\\

One of the major concerns in the field is the disparity in the results and fine details of the data reported by several groups. These can arise due to several parameters that strongly influence the electronic structure of the samples, such as twist angle, disorder, and dielectric environment.~A major challenge in the fabrication of small twist angle devices ($\theta<2\degree$) is that the atomic configurations are not thermodynamically stable ground states, and therefore tend to relax to a Bernal stacking order which has the minimum ground state energy.~In realistic samples, where a desired twist angle is targeted during the assembly of the heterostructures, an intrinsic strain is inevitable.~Such an uneven strain can lead to twist angle inhomogeneity in the sample.~Precise control of the twist angle is crucial, particularly in TBG, where the band structure is extremely sensitive to twist angles, and the robust correlated phases are found within the small range of magic angle.~Although correlated insulators are more robust and have been reproduced by many groups, superconductivity is found to be much more sensitive to microscopic variations of twist angle and disorder in the samples.~To date, no two devices have exhibited precisely the same superconducting response.~In addition, the results obtained by several groups show large variations in the critical temperature, critical magnetic field, critical current, and the density range over which the superconducting domes exist~\cite{cao2018unconventional, lu2019superconductors, saito2020independent, stepanov2020untying, arora2020superconductivity, liu2021tuning}.~The critical temperature $T_c$ and critical out-of-plane magnetic field $B_c$ are found to be maximum near $\theta=1.1\degree$~(Fig.~6f-g).~The measurements in some very high-quality samples (negligible twist angle disorder and clean interface) have revealed correlated insulators and superconductivity at or near all integer fillings of the flat bands~\cite{lu2019superconductors}.

\begin{figure*}[bth]
\includegraphics[width=0.9\textwidth]{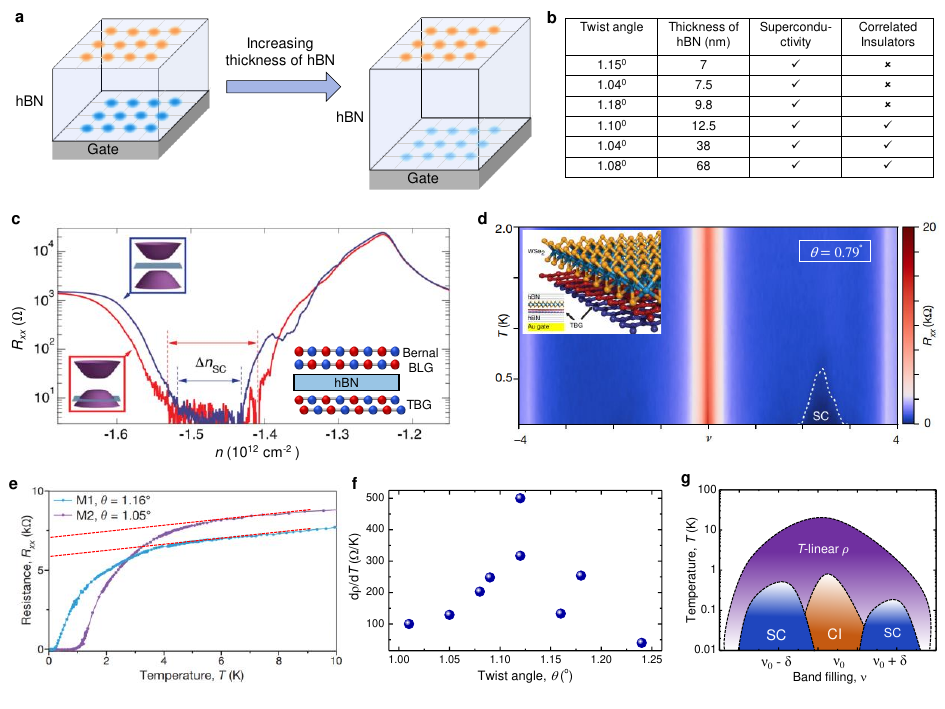}
\captionsetup{justification=raggedright,singlelinecheck=false}
\justify{\textbf{Fig.~7.~Tuning electronic correlations within the flat bands}.~\textbf{a.}~Schematic showing the charges on MATBG (orange) screened by the mirror charges (blue) on the graphite gate (gray). The graphite gate and TBG are separated by an hBN spacer (light blue). As the thickness of hBN increases, the strength of mirror charges on the graphite gate reduces, leading to stronger Coulomb repulsion in MATBG. In this way, Coulomb interactions can be effectively tuned in TBG.~\textbf{b.}~The table shows the devices with different twist angles and hBN thicknesses that have been investigated in ref.~\cite{stepanov2020untying, saito2020independent}. While superconductivity is robust for all the devices, correlated insulators are quenched for the first three devices with hBN thickness of 7 nm, 7.5 nm, and 9 nm, respectively.~\textbf{c.}~Fourprobe longitudinal resistance $R_{xx}$ measured as a function of density $n$ at a fixed displacement field $D$ for a MATBG in close proximity with a Bernal BLG. The BLG is charge neutral for the blue trace and hole-doped for the red trace. A broadening of superconductivity is observed when the Fermi energy is in the valence band of BLG. The schematic of the heterostructure is shown in the inset. Data taken with permission from ref.~\cite{liu2021tuning}. ~\textbf{d.}~$R_{xx}$ plotted as a function of filling factor $\nu$ and temperature $T$ for TBG proximitised by a monolayer tungsten diselenide (inset). Data taken with permission from ref.~\cite{arora2020superconductivity}. Here, a superconducting dome
outlined by a white dashed line has been observed for a twist angle of $0.79\degree$ which is much lower than the established range of magic angles. The absence of correlated insulators indicates that superconductivity and correlated insulators are governed by two different mechanisms.~\textbf{e.}~$R_{xx}$ as a function of $T$ near $\nu=-2$ for $\theta=1.16\degree$ and $1.05\degree$, taken with permission from ref.~\cite{cao2018unconventional}. Linear fits of these two curves above their superconducting transition temperatures give a slope of the order of $100~\Omega$/K.~\textbf{f.}~Estimated derivative of resistivity $\rho$ with respect to $T$, d$\rho$/d$T$ as a function of $\theta$ reported in ref.~\cite{cao2018unconventional, lu2019superconductors, saito2020independent, ghawri2022breakdown}.~d$\rho$/d$T$ lies in the range of $150-500~\Omega$/K for the magic angle regime.~\textbf{g.}~A schematic representation of the phase diagram seen near an integer filling $\nu_0$ in TBG. The correlated insulating (CI) state at $\nu_0$ is flanked by two superconducting (SC) domes at $\nu_0+\delta$ and $\nu_0-\delta$.~A metallic phase with $T$-linear resistivity is observed over a broad range of $\nu$, enclosing CI and SC~\cite{PhysRevLett.124.076801, jaoui2022quantum}.}
\end{figure*}

\subsection{\normalsize{Connection between correlated insulators and superconductivity}}

A straightforward approach to investigating the relationship between correlated insulators and superconductivity is to independently control their microscopic mechanisms. In MATBG, the presence of the flat bands leads to the condition $U/t\sim1$, where correlated insulators are not surprising. The first parameter, $t$, can be adjusted by altering the twist angle since the bandwidth is solely determined by the angle between the two graphene layers. The second parameter, $U$, can be independently controlled by engineering the dielectric environment.~The interactions between electrons in TBG can be manipulated via Coulomb screening by varying the dielectric separation between TBG and the metallic gate. This was demonstrated by fabricating multiple devices with varying thicknesses of the hBN gate dielectric~\cite{stepanov2020untying, saito2020independent}. The graphite screening layer, acting as the gate electrode, is expected to induce mirror charges in TBG (Fig.~7a). Consequently, when the distance between TBG and graphite gate is smaller than the moir\'{e} wavelength of approximately 13 nm, the polarised charge carriers screen out the Coulomb interactions and diminish the strength of $U$. A significant suppression of correlated insulators was observed in devices where the separation between TBG and the graphite gate was less than 12 nm, which is approximately the typical size of the Wannier orbital at the magic angle of $1.05\degree$ (see the table in Fig.~7b). Surprisingly, the suppression of correlated insulators does not affect superconductivity. Instead, superconductivity persists in all the samples tested, even for larger deviations from the magic angle. In a related work, the Coulomb interaction in MATBG was tuned by employing a Bernal BLG as a screening layer~\cite{liu2021tuning} (see the inset in Fig.~7c). In this case, a finite displacement field was utilised as a control knob to manipulate the band structure of BLG from metallic to insulating. When the Fermi energy was adjusted to lie within the band gap of BLG, there was no screening from BLG, and thus the insulating and superconducting phases in MATBG did not change significantly. However, when the Fermi energy was within the conduction or valence bands, the charge carriers in BLG could screen out the effective Coulomb potential in MATBG. Although the suppression of the correlated gap at $\nu=-2$ is observed, the state did not completely disappear, as observed in the previous studies~\cite{stepanov2020untying, saito2020independent} where hBN thickness was varied. Intriguingly, it was also found that the range of carrier densities over which the superconducting dome exists became broader (Fig.~7c), accompanied by an enhancement in the superconducting transition temperature. Overall, these findings indicate that correlated insulators and superconductivity exhibit opposite behaviours in the presence of Coulomb screening. Dielectric engineering was also explored by Arora et al. by introducing a layer of WSe$_2$ in proximity to TBG~\cite{arora2020superconductivity}.~Surprisingly, in this case, they observed superconductivity without correlated insulators at twist angles outside the established range of the magic angle (Fig.~7d).~All these reports suggest a disparate origin for these two phases, which compete with each other rather than being intimately connected. Consequently, these observations call into question the earlier interpretation of unconventional superconductivity, which was believed to be driven by electron-electron interactions.~Although twist angle variations and disorder can result in slight differences in the electronic structure of MATBG, engineering the dielectric environment presents new opportunities for studying a wider range of samples with greater diversity. This approach indeed holds the potential to shed light on the critical question surrounding the mechanism of superconductivity in MATBG as well as other multilayer graphene systems.

\subsection{\normalsize{Linear-in-temperature resistivity}}

One approach to finding clues about the origin of superconductivity involves studying the temperature dependence of resistivity in the normal metallic state, which has been previously explored in high $T_c$ superconductors~\cite{PhysRevLett.61.1658, RevModPhys.92.031001}.~A linear resistivity was observed in many cases, including cuprates~\cite{naqib2003temperature}, iron pnictides~\cite{PhysRevB.83.212506}, and ruthenates~\cite{PhysRevB.56.2916}.~Likewise, MATBG also exhibits a compelling linear relationship between resistivity and temperature near the correlated insulating states~\cite{cao2018unconventional, lu2019superconductors, saito2020independent, stepanov2020untying, liu2021tuning} (Fig.~7e). Deviation from the expected quadratic relationship between resistivity and temperature in a standard metal requires a compatible model beyond established mechanisms. In one of the early reports, linear resistivity was observed down to approximately $T=0.5$~K, but the presence of other correlated states hindered investigations at milliKelvin temperatures~\cite{PhysRevLett.124.076801}.\\

Our understanding of metallic phases is based on the behaviour of low-energy excitation within the periodic potential as Bloch waves. These interactions, involving all other quasiparticles, can be described using the Fermi liquid theory, which forms the foundation of the BCS theory of superconductivity~\cite{PhysRev.108.1175}.~For example, a state in a non-interacting system containing one particle with finite momentum transforms into an excited state with one quasiparticle of the same momentum. However, in correlated metals, the quasiparticle description may no longer hold due to strong electron interactions. These types of metals can be classified into two categories: bad metals (where the breakdown of quasiparticle description occurs at high temperatures) and strange metals (where it occurs at low temperatures)~\cite{doiron2003fermi, anderson2006strange, PhysRevB.105.235111}.~Strange metals possess unconventional electrical properties, such as linear-in-temperature resistivity and linear-in-field magnetoresistance.~In this regime, resistivity is governed by the scattering time scale in the Planckian limit $\tau_\hbar = \hbar/\alpha k_BT$, where $\hbar$ represents the reduced Planck constant, $k_B$ is Boltzmann's constant, and $\alpha$ is approximately of order unity.~The linear-in-$T$ resistivity observed in MATBG may indicate strange metallicity governed by strong electron-electron interactions for two reasons. First, the linear-in-temperature resistivity is observed over a wide range of filling factors near half-filling, suggesting a strong dependence on correlations within the flat bands~(Fig.~7g). Second, the presence of linear resistivity down to around $T=0.5$~K supports the likelihood of electron interactions rather than phonons. In a recent report, Coulomb screening from the graphite gates was used to suppress the correlated insulating states (as discussed in the previous section), which led to the observation of $T$-linear resistivity of the metallic state down to $T\rightarrow0$~\cite{jaoui2022quantum}. Additionally, it was also found that the resistivity varies linearly with the magnetic field, which is another characteristic of a strange metal. In another recent work, a strong deviation from the semiclassical Mott relation was observed in thermopower measurements of TBG over a range of twist angles in conjunction with linear-in-$T$ resistivity~\cite{ghawri2022breakdown}.~These signatures also alluded to a non-Fermi liquid behaviour and strange metallicity.\\

While these reports suggest a deviation from the conventional Fermi liquid theory, theoretical calculations proposed that electron-acoustic phonon scattering in TBG can lead to linear-in-temperature resistivity, which depends on the twist angle~\cite{PhysRevB.99.165112}.~At the magic angle, the reduced Fermi velocity results in a substantial increase in the electron-phonon coupling, resulting in an enhanced magnitude of resistivity. The slope in the resistivity-temperature data can reach a few hundred Ohm/Kelvin, consistent with experimental observations (Fig.~7f). Currently, arguments are proposed in favour of both strong electron-electron interactions and electron-phonon coupling. If the assumption of Fermi liquid theory breakdown holds true, MATBG could serve as an ideal platform to investigate the crucial relationship between superconductivity and strange metallic behaviour and may provide answers to several open questions in high-$T_c$ superconductors.
\begin{figure*}[bth]
\includegraphics[width=0.8\textwidth]{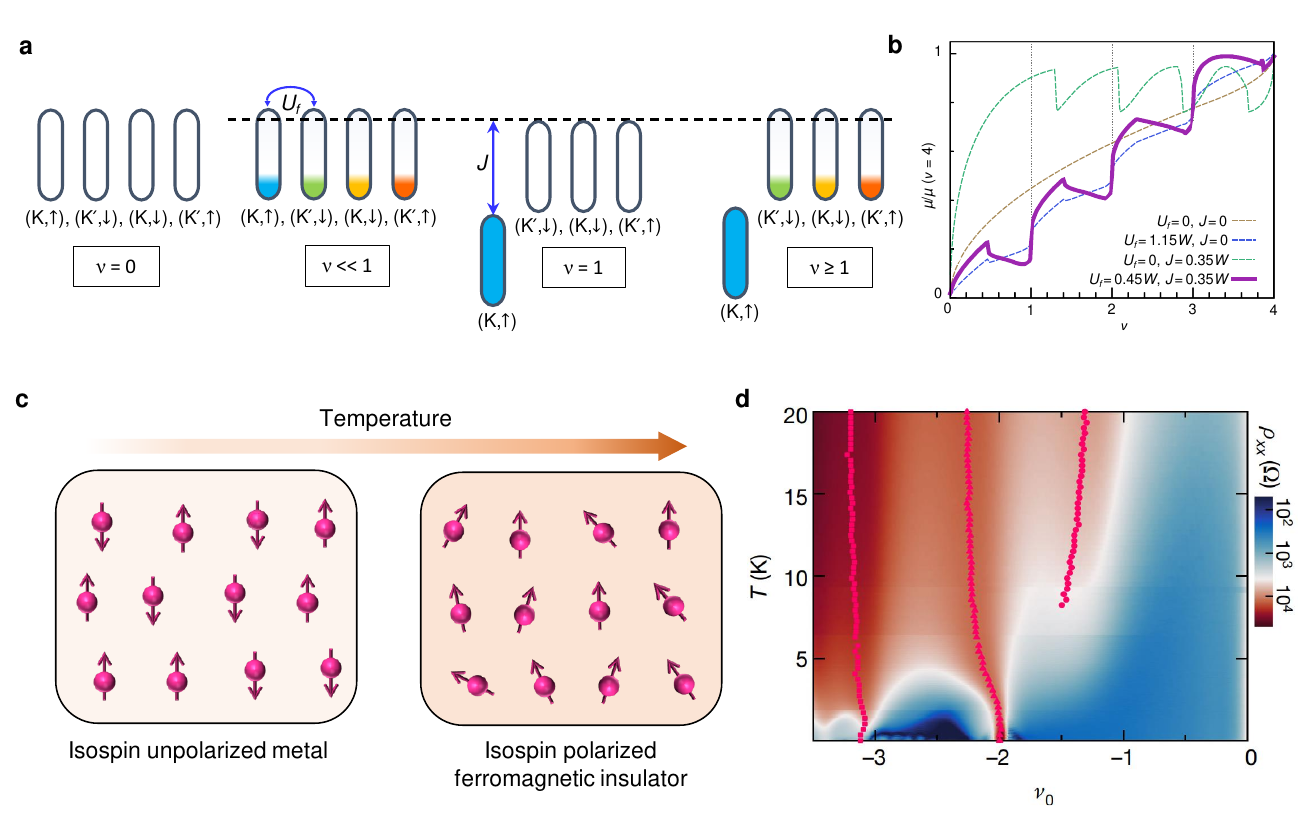}
\captionsetup{justification=raggedright,singlelinecheck=false}
\justify{\textbf{Fig.~8.~Flavours in twisted bilayer graphene}.~\textbf{a.}~Schematic of interaction-driven flavour symmetry breaking at $\nu=1$.~At $\nu<<1$, an inter-flavour Coulomb repulsion $U_f$ allows equal filling of each flavour illustrated by four different colours.~A phase transition takes place at $\nu=1$, where the system lowers its ground state energy by converting all the flavours into a single flavour which adds an intra-flavour exchange coupling $J$ in the Hamiltonian.~This process continues, and a similar phase transition occurs at all integer fillings.~\textbf{b.}~Mean field calculations of chemical potential $\mu$ as a function of $\nu$ for different values of $U_f$ and $J$, adapted with permission from ref.~\cite{park2021flavour}.~These results match perfectly with experimental data when both $U_f$ and $J$ are non-zero (solid purple curve).~\textbf{c.}~Illustration of the Pomeranchuk effect, where a transition from an isospin unpolarised metal to isospin polarised ferromagnetic insulator with low stiffness occurs when temperature increases.~\textbf{d.}~Resistivity $\rho_{xx}$ plotted as a function of $\nu$ and $T$, adapted with permission from ref.~\cite{saito2021isospin}.~At higher $T>5$~K, a resistive peak appears at $\nu=-1$, while no peak is seen at low temperatures.~The pink circles denote the position of local maxima in $\rho_{xx}$.}
\end{figure*}

\begin{figure*}[bth]
\includegraphics[width=0.8\textwidth]{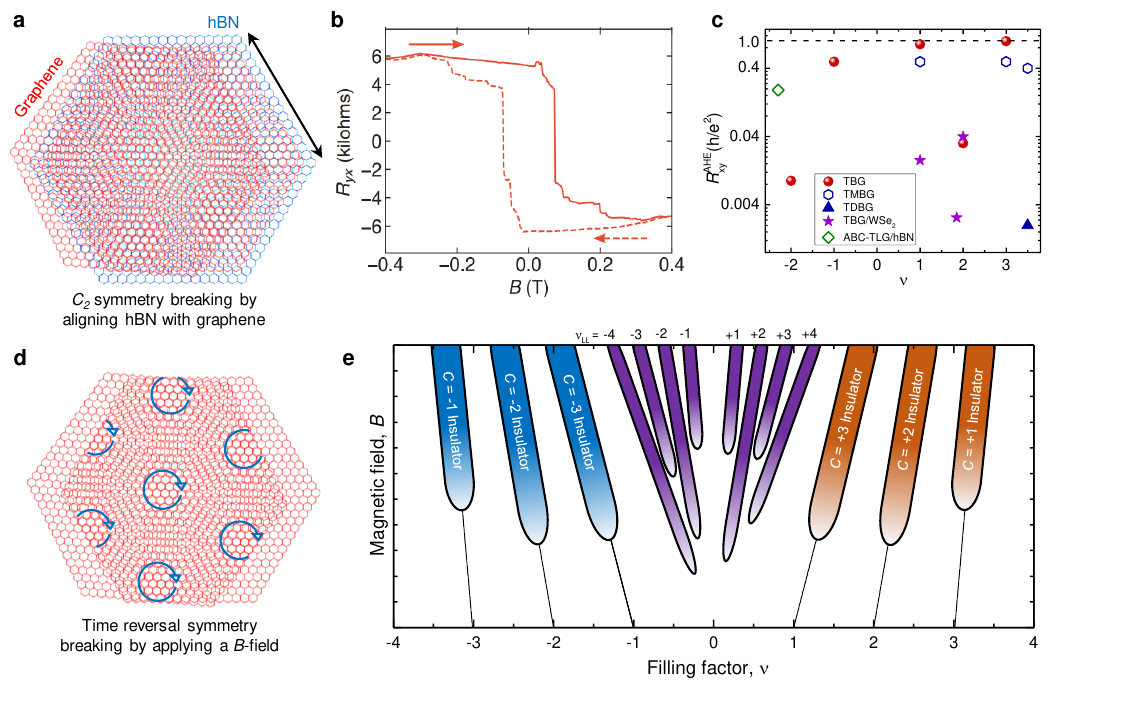}
\captionsetup{justification=raggedright,singlelinecheck=false}
\justify{\textbf{Fig.~9.~Topology of the flat bands}.~\textbf{a., d.}~The alignment of TBG with hBN breaks $C_2$ symmetry and application of a magnetic field $B$ breaks $\mathcal{T}$ symmetry.~The composite $C_2\mathcal{T}$ symmetry breaking opens up a gap at the Dirac points.~\textbf{b.}~Hall resistance $R_{xy}$ shows a hysteresis when the direction of magnetic field sweep changes indicating ferromagnetism.~This data is adapted with permission from ref.~\cite{sharpe2019emergent} which reported ferromagnetism near $\nu=3$ in TBG aligned with hBN.~\textbf{c.}~The plot shows the magnitude of $R_{xy}$ in units of $h/e^2$ at $B=0$ as reported by several groups~\cite{sharpe2019emergent, serlin2020intrinsic, PhysRevLett.127.197701, lu2019superconductors, tseng2022anomalous, polshyn2020electrical, polshyn2022topological, kuiri2022spontaneous, lin2022spin, bhowmik2023spin, chen2020tunable}.~Only a handful of reports shows quantised $R_{xy}$, while the other reports find $R_{xy}$ to be much lower than the expected quantised value.~\textbf{e.}~With the application of a finite $B$, additional Chern insulators emanate from different integer fillings.~In the phase space, these states are characterised by $R_{xx}\approx0$ and $R_{xy}=h/Ce^2$ where $C$ is the Chern number.~In addition, four-fold symmetry breaking gives rise to $8$ conventional quantum Hall states with Landau filling factors from $-4$ to $+4$, centered around $\nu=0$.}
\end{figure*}

\begin{figure*}[bth]
\includegraphics[width=0.8\textwidth]{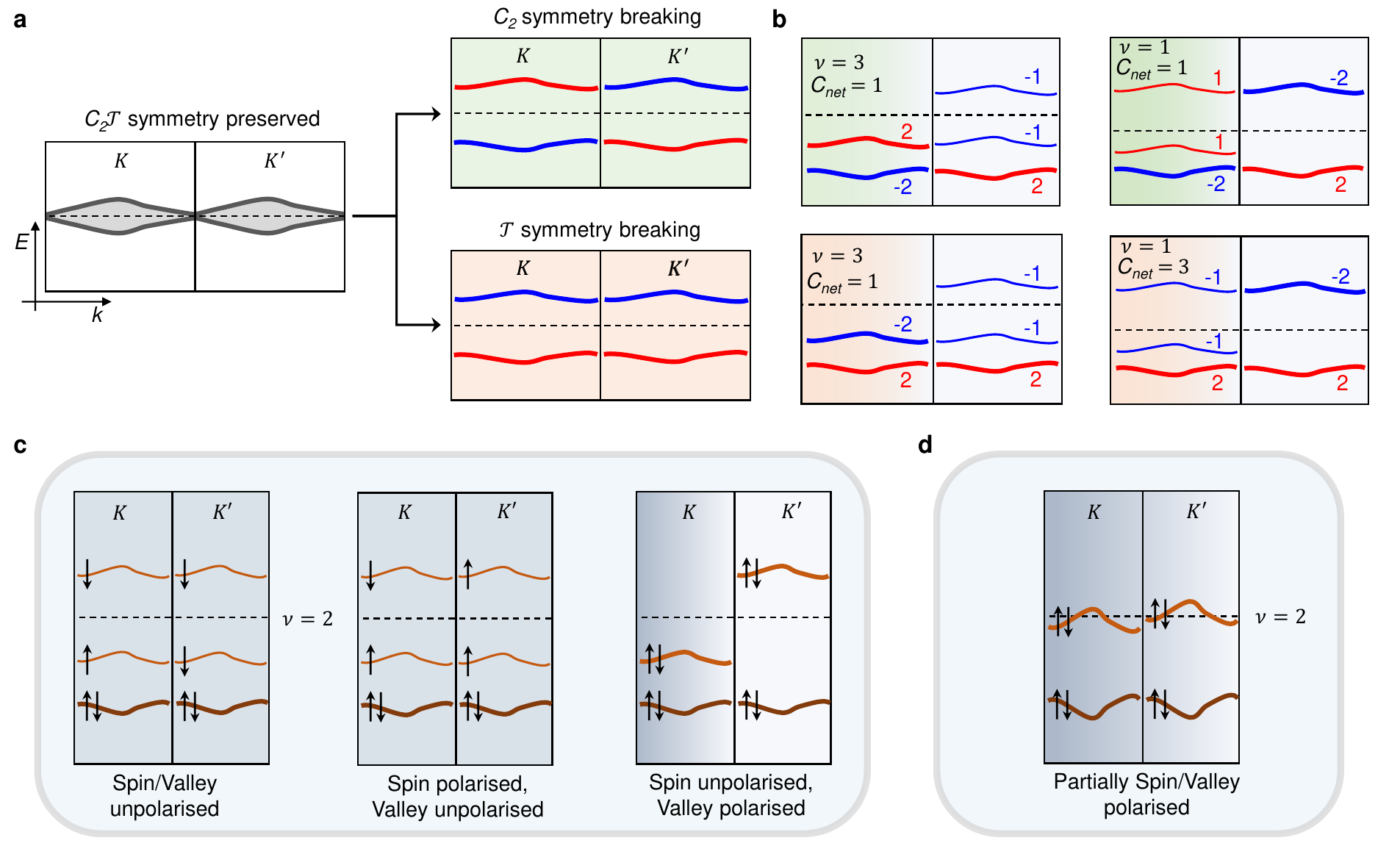}
\captionsetup{justification=raggedright,singlelinecheck=false}
\justify{\textbf{Fig.~10.~Schematic of the isolated Chern bands}.~\textbf{a.}~In TBG, the Dirac points at $K$ and $K^\prime$ valleys are protected by $C_2\mathcal{T}$ symmetry.~Breaking of $C_2$ symmetry lifts the degeneracy of the bands and creates isolated flat bands with opposite Chern numbers at $K$ and $K^\prime$ valleys.~Breaking of $\mathcal{T}$ symmetry opens up the same gap at both valleys leading to same valley Chern numbers.~There are a total of 8 bands (two spins, two valleys for conduction and valence bands) with each band possessing Chern number $C=+1(-1)$ denoted by red (blue).~\textbf{b.}~Sequential filling of these bands from $\nu=0$ results in different total $C$ at different $\nu$ for the two symmetry breaking cases. This is illustrated for $\nu=3$ and 1 for both $C_2$ (top green panels) and $\mathcal{T}$ symmetry breaking (bottom orange panels), respectively. The thicker and thinner bands denote spin degenerate and non-degenerate states, respectively.~These calculations of $C$ are in good agreement with experimental results.~\textbf{c.}~Schematic representing the three possible cases for $\nu=2$, with $C_2$ symmetry breaking. Valley polarised case, which leads to AHE, is shown as the third panel.~\textbf{d.}~Schematic showing a partially valley polarised scenario at $\nu=2$, where both $K$ and $K'$ valleys are partially occupied but with a net $K$ valley polarisation.}
\end{figure*}

\subsection{\normalsize Flavours in twisted bilayer graphene}

The electronic correlations in TBG are highly tunable with carrier density, and hence the partial filling of the low-energy flat bands can favour many different ground states. The presence of correlated insulators and superconductivity at or near all integer fillings raises questions about the role of Coulomb interactions at partial band fillings. While the correlated insulating states at $\nu=\pm2$ can be explained within the Hubbard model for a half-filled system, the states at $\nu=\pm1$ and $\pm3$ require an alternative model. In this section, we discuss how the sequential filling of the flat band, starting from the CNP, governs the correlated phases observed in experiments.\\

The non-dispersive electronic bands in TBG are four-fold degenerate due to spin and valley degrees of freedom, with each commonly referred to as an `isospin flavour'. To explain the correlated gap opening at the integer fillings, a toy model is widely used~\cite{zondiner2020cascade, park2021flavour, paul2022interaction, saito2021isospin, wong2020cascade}~(Fig.~8a).~The model assumes that when electrons are added to the system, all the flavours are equally occupied at very low densities near the CNP, and the ground state is determined by the inter-flavour Coulomb interactions $U_f$. As the density increases toward $\nu=\pm1$, an intra-flavour Hund's coupling $J$ comes into play and competes with $U_f$.~At $\nu=\pm1$, the dominance of $J$ over $U_f$ lifts the isospin degeneracy, causing all four flavours to merge into a single flavour.~Indeed, the breaking of isospin-symmetry has two significant outcomes: the opening of a correlated gap at $\nu=\pm1$ and the revival of Dirac-like behaviour for the three remaining unoccupied flavours. This process continues, leading to a cascade of phase transitions at all integer fillings. These unique band transformations at integer fillings have been observed in experiments using a nanotube-based scanning single electron transistor, which measures electronic compressibility in MATBG~\cite{zondiner2020cascade}.~Fig.~8b presents theoretical calculations of the chemical potential $\mu$ as a function of $\nu$ for different values of $U_f$ and $J$~\cite{park2021flavour}.~As the flavours fill starting from $\nu=0$, $\mu$ initially increases and reaches a local maximum around $\nu=0.6$. Surprisingly, $\mu$ sharply decreases, exhibiting negative inverse compressibility d$\mu$/d$n$, and stabilises at a local minimum at $\nu\approx1$. This process repeats at each integer filling. Negative compressibility indicates an incompressible state and corresponds to the opening of correlated gaps at partial band fillings. Overall, this model aligns well with the observation of correlated insulators at all integer fillings in MATBG.\\

The correlated insulating states discussed above are usually found at low temperatures, typically below $T=40$ K. Thermodynamically, higher temperatures favour states with higher entropy, leading to the appearance of metallic states at higher temperatures. Likewise, insulating states with ordered electrons and lower entropy are typically stabilised at low temperatures. In many condensed matter systems, metal-to-insulator transitions occur when temperatures are lowered. For instance, the correlated phases in MATBG are expected to emerge at low temperatures.~Intriguingly, Saito et al. and Rozen et al. reported emergence of insulating states at a higher temperature, near $\nu=\pm1$~\cite{saito2021isospin, rozen2021entropic}.~In order to explain the possible mechanism, a close analogy was drawn between MATBG and the `Pomeranchuk effect' in Helium-3, which becomes a solid at a higher temperature~\cite{RevModPhys.69.683}~(Fig.~8c).~This phenomenon is thermodynamically feasible because solid Helium-3 has higher entropy than its liquid form due to the spin fluctuations of Helium-3 atoms at higher temperatures. In MATBG, the transition from metal to insulator above $T\approx 5-10$ K suggests a Pomeranchuk-like mechanism (Fig.~8d), with the insulating state persisting up to $T\approx70-100$ K. By measuring the system's entropy, it was found that the entropy of the high-temperature insulating phase is higher than that of the low-temperature metallic phase, which can be explained in terms of spin-valley isospin fluctuations in MATBG. At low temperatures, when the system behaves as a metal, it is isospin unpolarised, meaning it has an equal number of isospins in two opposite directions, resulting in lower entropy due to their ordering. However, at high temperatures, the system acts as a ferromagnetic insulator, with flavours broadly aligned in a particular direction but with low isospin stiffness, leading to finite isospin polarisation at $\nu=\pm1$. To gain a better understanding of the phase transition from an unpolarised to a polarised state at higher temperatures, future experiments in MATBG, along with other moir\'{e} systems, are necessary. Specifically, investigations conducted in the presence of an out-of-plane magnetic field would be valuable to explore the possible ferromagnetic ordering at higher temperatures.

\subsection{\normalsize Topological Flat Bands}

Having discussed correlated insulators, superconductivity, and their mutual connection, we now turn our attention to another crucial aspect of moir\'{e} graphene systems, namely, topology. In this section, we present an overview of the emergence of novel electronic phases resulting from the interplay between interactions and the topology of flat bands. In condensed matter physics, topology finds its significance mainly in the context of the quantum Hall effect for 2D electron gases subjected to a strong out-of-plane magnetic field $B$~\cite{tong2016quantum, RevModPhys.71.298, bolotin2009observation}.~Under these conditions, the energy levels of the electrons are Landau quantised, with a gap between the levels proportional to $B$. While electrons move in cyclotron orbits in the bulk of the system, a distinct behaviour arises at the system's edges, where the hard-wall boundary conditions do not allow electrons to escape from the system. Consequently, electrons follow half of a cyclotron orbit, resulting in unidirectional skipping orbits along the edges. These states, squeezed at the sample's boundary, are known as edge states and are crucial for realising the quantum Hall effect. When the chemical potential lies between two Landau levels, transport occurs primarily through the edge modes, leading to a quantised transverse conductivity of $Ne^2/h$, where $N$ is the filling factor representing the number of edge states or the total number of occupied Landau levels. The filling factor $N$ is a topological invariant that remains robust to many external perturbations.~An example of a topological invariant is the winding number, which only depends on the winding around a point and not on the specific winding path or how the winding is done~\cite{PhysRevD.46.5607}.~This concept is particularly relevant when considering chiral electrons' current loops, which generate orbital magnetic moments. The strong $B$-field reduces the kinetic energy of the electrons, resulting in perfectly flat Landau levels where the Coulomb energy dominates over the electrons' kinetic energy.~In contrast, MATBG possesses a set of low-energy flat bands even at zero $B$, making it an ideal candidate for exploring the interplay between interactions and topology. We refer the readers to reference \cite{tong2016quantum} for detailed discussions on topology and quantum Hall effect in 2D electron systems.\\

Several theoretical studies have explored the flat band topology in TBG~\cite{PhysRevLett.123.036401, PhysRevLett.124.167002, PhysRevLett.123.216803, PhysRevLett.129.047601, PhysRevB.99.035111, PhysRevX.9.021013, PhysRevX.11.011014, PhysRevLett.124.187601}.~The low-energy bands in TBG can be described as 2D Dirac cones in the presence of a pseudo magnetic field generated by the moir\'{e} pattern~\cite{PhysRevB.99.155415}.~It has been shown that the bands are topologically non-trivial since they can be characterised by Wilson loops with odd winding numbers~\cite{doi:10.1021/acs.nanolett.2c01481, PhysRevX.9.021013, PhysRevB.103.035427}.~Such non-trivial topology arises from the fact that the two flat bands for each valley are the zeroth Landau levels of Dirac fermions coupled to opposite pseudo magnetic fields leading to opposite sublattice polarisation. The topology of a 2D system is determined by the integration of Berry curvature over the first Brillouin zone defined as Chern number $C$, which becomes $\pm1$ for the filled bands of TBG~\cite{liu2021orbital}.~Overall, a total eight flat bands in TBG (conduction and valence bands with two spins and two valleys) can be viewed as eight zeroth Landau levels of Dirac fermions and the degeneracy is protected by a composite $C_2\mathcal{T}$ symmetry.~It has been predicted theoretically that the spin-valley degeneracy of these eight bands can be lifted by strong Coulomb interactions, leading to fully spin-valley polarised insulating states with finite Chern numbers at the integer band fillings~\cite{PhysRevLett.123.036401, PhysRevLett.124.167002, PhysRevLett.123.216803, PhysRevB.103.035427}.~The $C_2$ symmetry in TBG can be broken by aligning with a layer of hBN, while the application of a magnetic field breaks $\mathcal{T}$ symmetry~\cite{PhysRevB.103.075122, PhysRevResearch.1.033126, PhysRevResearch.3.013242, PhysRevB.102.081118, PhysRevB.107.064501, PhysRevX.10.031034}~(Fig.~9a,~9d).~A natural consequence of Chern bands is the realisation of quantised transverse conductance, even when the external magnetic field is zero.~The first observation of an anomalous Hall effect (AHE) was reported at $\nu=3$ in a MATBG aligned with hBN~\cite{sharpe2019emergent}~(Fig.~9b).~A hysteresis was observed in the transverse resistance with respect to an out-of-plane magnetic field, suggesting ferromagnetism.~According to the theoretical calculations breaking of $C_2$ symmetry results in a gapped state with $C=\pm1$ indicative of quantum anomalous Hall insulators at $\nu=\pm3$ with a quantised transverse resistance of $R_{xy}=\pm h/e^2$~\cite{PhysRevLett.124.166601}.~However, the observed maximum $R_{xy}$ of $10.4$~k$\ohm$ was lower than the quantised value.~Later a quantised $R_{xy}=h/e^2$ at $\nu=3$ in hBN-aligned MATBG was observed in a different study~\cite{serlin2020intrinsic}.~This disparity among the samples has been a major limitation in realising quantum AHE, also called `Chern insulators', in TBG and to date, there is only a handful of reports, spanning different moir\'{e} systems and band fillings~\cite{serlin2020intrinsic, PhysRevLett.127.197701, doi:10.1021/acs.nanolett.1c00696, tseng2022anomalous, lin2022spin, bhowmik2023spin, polshyn2020electrical, polshyn2022topological, he2021competing, chen2021electrically, he2021chirality, kuiri2022spontaneous, chen2020tunable, chen2022tunable}.~In Fig.~9c, we have presented a graph that shows the magnitude of $R_{xy}$ at $B=0$ seen at different $\nu$ for various moir\'{e} graphene systems.~It is clear from the plot that in most of the cases the reported $R_{xy}$ is much lower the expected quantised value.~As discussed earlier, twist angle inhomogeneity and strain obscure the reproducibility of data in moir\'{e} graphene and no two devices have exhibited exactly similar characteristics.~Particularly, at $\nu=\pm1$ and $\pm3$, where fully spin-valley polarised quantum anomalous Hall insulators are expected, twist angle disorder and domain walls can produce conducting bulk modes parallel to the edge channel leading to a reduction in anomalous $R_{xy}$.~In a recent experiment utilising a scanning superconducting quantum interference device (SQUID), measurement of local magnetisation was performed in MATBG~\cite{grover2022chern}.~This technique has revealed multiple magnetic domains which reverse upon flipping the direction of the density sweep. Hence, in electrical transport measurements, such reversible magnetic moments do not result in hysteresis, despite the existence of local magnetic moments within the sample.~These findings provide further insights into the lack of reproducibility of ferromagnetism in TBG samples, where hysteresis is the only indicator of magnetism.~While the realisation of quantised Hall insulators remains a challenge due to the aforementioned limitations, the observation of AHE in these systems itself is novel and arises solely from the combined effect of interactions and band topology.\\

Next, we focus on the ground states of these ferromagnets.~A conventional ferromagnet can have spin and orbital components, and depending on their relative strength, a ferromagnet can be classified into two categories: spin ferromagnet and orbital ferromagnet.~The origin of magnetism in TBG is purely orbital since a weak SOC in graphene is not sufficient to give rise to a spin ferromagnet.~Although it was known that the loop currents carried by electrons around $K$ and $K^\prime$ valleys in graphene give rise to an intrinsic orbital magnetic moment, the net effect was found to be zero since the $K$ and $K^\prime$ valleys have opposite moments that cancel each other~\cite{PhysRevLett.99.236809}.~However, in TBG, a net valley polarisation is possible and several experimental signatures discussed below support the orbital nature of magnetism. It was theoretically proposed that valley polarisation leading to ferromagnetism can flip upon changing the doping in an orbital Chern insulator when the Fermi energy is below the Chern gap~\cite{PhysRevLett.125.227702}.~Such a reversal of magnetisation observed in TMBG at $\nu=3$~\cite{polshyn2020electrical} and TBG at $\nu=2$~\cite{bhowmik2023spin} indicate valley-polarised ground states. In addition, a large magnetic moment of the order of one Bohr magneton measured near $\nu=3$ in TBG~\cite{tschirhart2021imaging} is expected only if the chiral edge mode contributions of an orbital Chern insulator are considered.~Subsequent transport measurements conducted on TBG have provided evidence that the magnetism in the system exhibits a high degree of anisotropy~\cite{doi:10.1021/acs.nanolett.1c00696}.~This observation was reported by performing measurements at various angles while applying both out-of-plane and in-plane magnetic fields. When a higher out-of-plane field and a smaller in-plane field are applied, the hysteretic behaviour remains consistent with that observed in the presence of a purely out-of-plane magnetic field. However, as the in-plane field strength is increased, the hysteresis becomes weaker, and at the highest in-plane field, the hysteretic transitions in $R_{xy}$ reduce significantly. The results suggest that the dominant factor governing magnetism in TBG is the valley (orbital) degree of freedom. The anisotropic behaviour of magnetism further supports the notion that valleys, rather than isotropic spins, are the leading source of magnetism in TBG.\\

It is worth noting that the band fillings can have a substantial impact on stabilising magnetism in TBG since the number of spin/valley flavours and their polarisation depends on the carrier density in the system. The $C_2\mathcal{T}$-symmetry broken ground state at even integer fillings ($\nu=\pm2$) is degenerate, unlike the odd integer fillings ($\nu=\pm1,\pm3$). At $\nu=\pm2$, three degenerate ground states are: \textbf{1.} spin/valley-unpolarised, \textbf{2.} spin-polarised and valley-unpolarised, and \textbf{3.} spin-unpolarised and valley-polarised (Fig.~10c). The degeneracy among these states is lifted by an inter-valley Hund's coupling, favouring either of the first two states~\cite{PhysRevLett.124.187601, PhysRevLett.124.166601}. There is no valley polarisation in that case, and hence, ferromagnetism is unexpected at $\nu=2$. The scenario can change significantly in the presence of SOC, which can be induced in TBG by partnering it with a layer of WSe$_2$. The proximity-induced SOC can break $C_2$ symmetry and surpass the effect of inter-valley Hund's coupling by introducing an additional energy term in the Hamiltonian. Therefore, a combination of $C_2$ symmetry breaking and SOC can lead to a finite valley polarisation, favouring a ferromagnetic ground state at $\nu=\pm2$ as observed experimentally in TBG/WSe$_2$ systems~\cite{lin2022spin, bhowmik2023spin}.~The reversal of magnetisation controlled by carrier density is considered a strong indicator of valley ordering and is a tell-tale signature of orbital ferromagnetism~\cite{bhowmik2023spin}. The presence of SOC can promote a coupling between spin and valley order, as evidenced by the observed hysteresis with respect to an in-plane $B$-field~\cite{lin2022spin}. The coupling between the in-plane field and magnetic order arises from the combined effect of SOC and spin Zeeman energy. While SOC is considered a crucial factor in realising AHE at $\nu=2$, other factors, such as the competition between Coulomb repulsion and increasing band dispersion away from the magic angle, may lead to partially valley-polarised states at $\nu=2$ (Fig.~10d). Recent experiments have found AHE at $\nu=\pm2$ in TBG aligned with hBN~\cite{tseng2022anomalous}. Similar observations in TBG/WSe$_2$ and hBN-aligned TBG suggest that the valleys in these systems might polarise through different mechanisms, and the precise nature of magnetism at $\nu=\pm2$ remains an open question. Additionally, apart from TBG, other 2D moir\'{e} materials such as ABC-trilayer graphene aligned with hBN~\cite{chen2020tunable, chen2022tunable}, TMBG~\cite{polshyn2020electrical, polshyn2022topological, chen2021electrically}, and MoTe$_2$/WSe$_2$~\cite{tschirhart2023intrinsic} exhibit similar ferromagnetic signatures that are believed to be orbital.\\



Next, we discuss the effects of $\mathcal{T}$-symmetry breaking via the application of a modest $B-$field.~Scanning tunneling microscopy in MATBG without hBN alignment first revealed a sequence of topological Chern insulators stabilised by a perpendicular $B$-field~\cite{nuckolls2020strongly}.~Such insulators at $\nu=\pm1,\pm2,\pm3$ are characterised by $C=\pm3,\pm2,\pm1$ following Diophantine equation, $n/n_0 = C\phi/\phi_0 + s$, where $n_0$ is the density corresponding to one carrier per moir\'{e} unit cell, $\phi$ is the magnetic flux per moir\'{e} unit cell, $\phi_0 = h/e$ is the flux-quantum, $h$ is Planck's constant, $e$ is the charge of the electron, and in the context of Chern insulators the band filling index $\nu$ at $B = 0$ T is commonly referred to as $s$~(Fig.~9e).~Notably, this sequence was found to be incompatible with the previously predicted Chern numbers for the $C_2$ symmetry breaking scenario.~In particular, when $C_2$ symmetry is broken, $C=\pm1$ is expected at $\nu=\pm1$ which is inconsistent with the observation of $C=\pm3$  (Fig.~10b).~This scenario is explained by $\mathcal{T}$ symmetry breaking via strong electronic interactions that produce Chern bands with equal Chern numbers of $C=+1(-1)$ in the $K$ and $K^\prime$ valleys.~Strong correlations can break $\mathcal{T}$ symmetry even at $B=0$ stabilising these states near to zero-$B$~\cite{PhysRevB.102.081118}.~In Fig.~10a-b, we have presented a schematic of the Chern bands produced when either $C_2$ or $\mathcal{T}$ symmetry is broken, and they can independently open an energy gap with opposite signs at the Dirac point.~The sequential filling of these isolated bands produces spin-valley polarised insulating states at integer fillings consistent with the experimental observations~\cite{nuckolls2020strongly, wu2021chern, das2021symmetry, saito2021hofstadter, bhowmik2023spin}.

\begin{figure*}[bth]
\includegraphics[width=0.8\textwidth]{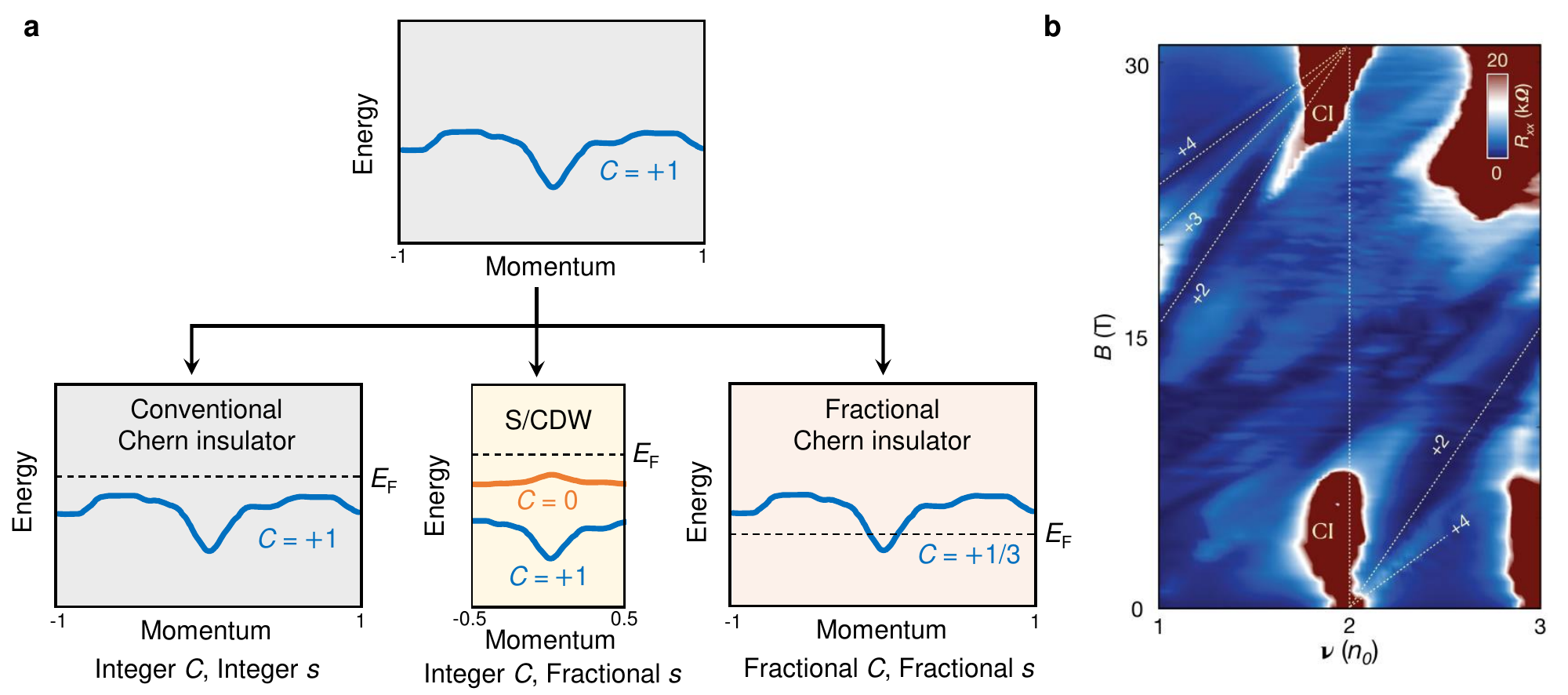}
\captionsetup{justification=raggedright,singlelinecheck=false}
\justify{\textbf{Fig.~11.~Unconventional quantum Hall phases}.~\textbf{a.}~Based on the experimental observations, the unconventional quantum Hall insulators are classified into three categories: conventional Chern insulator (integer $C$, integer $s$), spin/charge density wave (integer $C$, fractional $s$), and fractional Chern insulator (fractional $C$, fractional $s$).~We call $C_2\mathcal{T}$ broken Chern insulator the conventional Chern insulator.~When translation symmetry is broken the unit cell becomes a few times larger than original unit cell depending on the order of symmetry breaking.~Hence, the Brillouin zone becomes folded leading to additional Chern bands in the system.~For instance, for a unit cell doubling, $C=+1$ band splits into $C=+1$ and $C=0$, with a Brillouin zone that is halved~\cite{pierce2021unconventional, polshyn2022topological}.~Fractional filling of these bands at an intermediate field leads to a phase transition from charge density wave to fractional Chern insulator~\cite{xie2021fractional}.~\textbf{b.}~Magnetoresistance $R_{xx}$ plotted as a function of $\nu$ and $B$ adapted with permission from ref.~\cite{PhysRevLett.128.217701}.~The correlated insulating state at $\nu=2$ first disappears with increasing $B$ and re-enters at a sufficiently high $B>25$~T.~While the Landau levels at low field point towards higher density from $\nu=2$, at the high field this scenario reverses suggesting two different mechanisms leading to these two phases in low and high $B$-fields.}
\end{figure*}

\subsection{\normalsize Unconventional Quantum Hall phases}

Aside from the non-trivial insulating states stabilised by electron-electron interactions, other exotic phases have also been observed in TBG, which cannot be explained by considering $C_2\mathcal{T}$ symmetry breaking alone. In this section, we will provide an overview of some of these exotic phases.~The discovery of Chern insulators at integer band fillings in TBG naturally raises the possibility of realising these states at fractional fillings and/or the possibility to stabilise fractional Chern insulators.~While interacting lattice models with flat Chern bands have suggested the possibility of fractional quantum Hall states at zero $B$~\cite{PhysRevB.99.075127, PhysRevB.86.201101, PhysRevLett.109.186805, PhysRevLett.116.216802}, determining the experimental conditions for probing these phenomena has been a significant challenge.~Flat bands in TBG provide an ideal platform to explore these states, particularly in samples featuring low disorder and large-scale homogeneity. 
~The possible insulating states can be classified into three different classes based on the Chern number $C$ and band filling index $s$: \textbf{a.}~Conventional Chern insulators ($C_2\mathcal{T}$ broken, integer $C$, integer $s$),~\textbf{b.}~Spin/Charge density wave (S/CDW) states (Translation symmetry broken, integer $C$, fractional $s$),~\textbf{c.}~Fractional Chern insulators (fractional $C$, fractional $s$)~(Fig.~11a).~In recent reports of electronic compressibility measurements~\cite{pierce2021unconventional, xie2021fractional}, while an abundance of new states with negative compressibility at different integer fillings of the bands, indicated the presence of conventional Chern insulators, additional states with fractional $C$ and/or $s$ were also observed. The unusual sequence of Chern insulators at fractional fillings, and their existence down to zero $B$-field suggests the possibility of topological S/CDW order in the system. The occurrence of fractional fillings in TBG can be attributed to the existence of quasiparticle excitations with fractional charge ($e/f$, where $f$ determines the order of the fraction), which is responsible for the fractional quantum Hall effect. Most importantly, in that case, the Hall resistance is also expected to be fractionally quantised. However, it is found that some of the Chern insulators at fractional fillings persisting down to $B=0$ do not exhibit fractional Chern numbers.~Therefore, in such a case, the underlying ground state is inferred to have a density ordering that breaks the translation symmetry in the system, resulting in an enlarged moir\'{e} unit cell.~In this scenario, an integer number of electrons can occupy the ground state of this new unit cell, causing the number of electrons per original unit cell to become fractional.~In Fig.~11a, we present how the doubling of the moir\'{e} unit cell, favoured by S/CDW, can lower the system's ground state energy by generating new Chern bands.~For example, in a doubled unit cell, the $C=+1$ band can split into two bands with $C=0$ and $C=+1$.~The modified $C=+1$ band has lower energy compared to the original $C=+1$ band and a new Chern band, $C=0$ is formed.~Recently, transport measurements in TMBG~\cite{polshyn2022topological} and TBG/WSe$_2$~\cite{bhowmik2022broken} in the presence of a perpendicular $B$ field have uncovered such new states at half-integer fillings.~In TMBG, new Chern insulating states with $C=1$ emerge from $\nu=1.5$ and $3.5$ in the zero $B$-field limit.~Similarly, in TBG/WSe$_2$, the observation of Lifshitz transitions at $\nu=0.5$ and $3.5$ was found to be consistent with a doubled moir\'{e} unit cell.~In the latter report, these states have also been detected in thermoelectric transport measurements even at zero $B$-field, suggesting the presence of a S/CDW in the system~\cite{bhowmik2022broken}.~Despite the experimental evidence pointing towards a translation symmetry-broken lattice, the microscopic origin of the S/CDW order in these systems remains elusive.\\

Fractional statistics in the quantum Hall limit is expected at a large $B$-field and has been observed in a very few systems such as Bernal BLG aligned with hBN~\cite{spanton2018observation}.~By contrast, the topological flat bands in MATBG favour such states at zero $B$-field.~While fractionally filled states with integer Chern numbers require a S/CDW order, fractional Chern insulators are compatible with quasiparticle excitations with fractional charge. Unlike fractional quantum Hall insulators, they are stabilised near zero $B$-field, indicating the non-trivial topology of the flat bands. In Fig.~11a, we illustrate the schematic of a fractional Chern insulator with $C=1/3,~s=1/3$, observed in reference~\cite{xie2021fractional}.~It is interesting to note that typically these unconventional quantum Hall insulators are in competition with each other.~A phase transition from a CDW-like ground state to fractional Chern insulators is anticipated as the magnetic field is increased and the ratio of inter- and intra-sublattice coupling changes~\cite{pierce2021unconventional, xie2021fractional}.~However, controlling this ratio in experiments remains a significant challenge.\\

To date, most of the experiments have focused on electron transport in zero and intermediate (usually below $B=10$~T) $B$-field limit.~However, the single particle bandstructure in the non-interacting regime suggests Hofstadter subbands at high$-B$ in MATBG, which was previously unexplored.~Recently, Das et al.~\cite{PhysRevLett.128.217701} studied the magnetotransport in MATBG upto $B=31$~T which produces one magnetic flux quantum per moir\'{e} unit cell.~In addition to Hofstadter butterfly, they also found a re-entrant correlated insulator at half filling $\nu=+2$ (Fig.~11b).~Theoretical analysis predicts a new set of flat bands at one magnetic flux quantum leading to reappearance of correlated insulating state at $\nu=+2$~\cite{PhysRevLett.129.076401}.~These bands are different from those at $B=0$ in terms of symmetries and topology and will require follow-up theory and experiments for better understanding.\\

\begin{figure*}[tb!]
\includegraphics[width=0.75\textwidth]{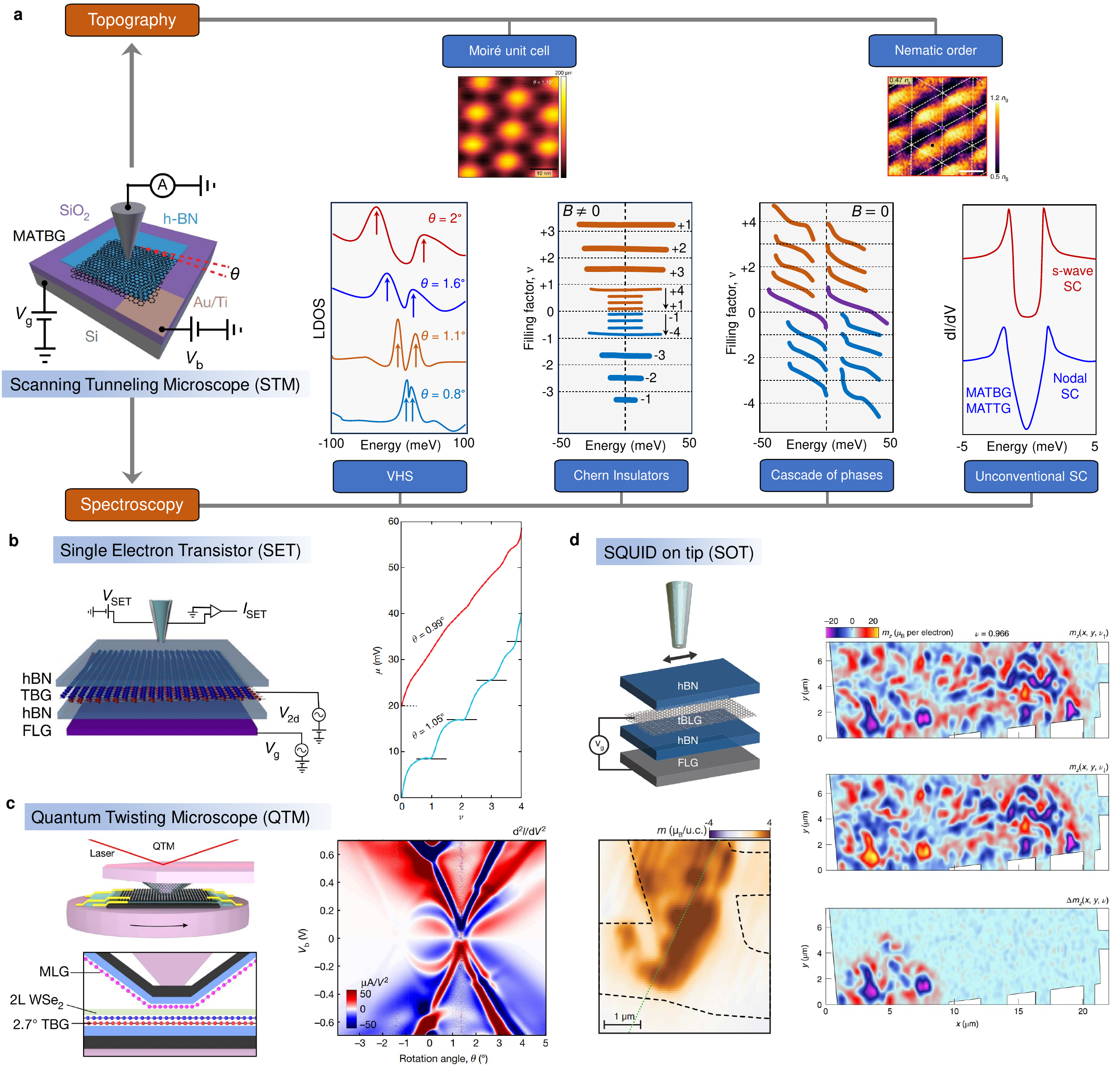}
\captionsetup{justification=raggedright,singlelinecheck=false}
\justify{\textbf{Fig.~12.~Scanning probe techniques in magic-angle twisted bilayer graphene.}~\textbf{a.}~Schematic of the STM setup, where MATBG is placed below a tungsten (W) tip. Measurements using STM fall under two broad categories: topography and spectroscopy. In the topography section (upper panel), the spatial mapping of the height distribution shows AA (bright) and AB (dark) sites in the MATBG sample (data taken with permission from ref.~\cite{kerelsky2019maximized}). The local density profile near the half-filling of the band shows stripe-like features indicating the existence of a nematic order (data taken with permission from ref.~\cite{rubio2022moire}).~The bottom panel shows schematics of several spectroscopic features in MATBG investigated by STS. The measurement of the local density of states (LDOS) d$I$/d$V$ is performed by applying a bias voltage $V_b$, and carrier density is tuned using a global Si or graphite bottom gate.~The first plot shows LDOS measured at zero doping on AA sites for four different twist angles 2$\degree$-0.8$\degree$ (right). As the angle is reduced, the energy separation between two peaks indicative of vHSs decreases. The second plot shows the schematic representation of the Chern insulators observed at a finite magnetic field. The states in the conduction and valence bands are denoted by orange and blue colours, respectively. The third plot shows a cascade of phase transitions observed at different fillings in the conduction (orange) and valence (blue) bands. The fourth plot shows the d$I$/d$V$ profile for a s-wave conventional superconductor (U-shaped) and a nodal superconductor (V-shaped). Both MATBG and MATTG have shown nodal superconducting nature.~\textbf{b.}~Schematic of the SET setup~\cite{yu2022correlated} (left). Chemical potential $\mu$ vs. filling $\nu$ measured for two different twist angles of 0.99$\degree$ and 1.05$\degree$ shows pinning of chemical potential at all integer $\nu$, further indicating identical phase transitions (data taken with permission from ref.~\cite{zondiner2020cascade}).~\textbf{c.}~Schematic of the QTM setup~\cite{inbar2023quantum} (top left). An AFM tip with a motorised rotator is brought into contact with the sample, providing in-situ control over the twist angle. Momentum-resolved tunneling is demonstrated using a device composed of monolayer graphene and TBG separated by a WSe$_2$ barrier (bottom left). The second derivative of d$I$/d$V$ as a function of $V_b$ and rotation angle measured in tunnel device of TBG (right). The colour plot clearly shows the low-energy flat and high-energy remote bands of TBG (data taken with permission from ref.~\cite{inbar2023quantum}).~\textbf{d.}~Schematic of the nanoSQUID on tip setup (top left). Imaging of local magnetism over some portion of the total area of the sample where the black dashed lines indicate the edge of the sample~\cite{tschirhart2021imaging} (bottom left). Data shows the magnetisation density in units of Bohr magnetons per moir\'{e} unit cell~\cite{tschirhart2021imaging}. In data taken from a different work~\cite{grover2022chern} (three vertical panels on the right), magnetisation density is presented for two opposite directions of density sweeps near $\nu=1$ showing local magnetic moments, most of which are reversible upon reversing the direction of the sweep. Therefore, the net magnetisation (third plot) remains non-zero only over a tiny region in the sample.}
\end{figure*}

\section{Scanning Techniques to probe electronic correlations}

Electrical transport measurements in van der Waals heterostructures are often limited by the lateral dimensions of the devices, typically of the order of a few micrometers. However, in moir\' {e} systems, electronic correlations become significant over superlattice periods of a few tens of nanometers. Consequently, the transport measurements discussed so far probe various global, averaged quantities. In recent years, local scanning probes capable of capturing electron-electron interactions with atomic precision have been employed that provide deeper insights into the intricacies of moir\' {e} systems. This section focuses on four scanning probe tools, namely, scanning tunneling microscope (STM), single-electron transistor (SET), quantum twisting microscope (QTM), and nano-SQUID on the tip (SOT).\\

The working principles of STM, SET, and QTM rely on the quantum tunneling of electrons between a conducting tip and the sample~\cite{binnig1983scanning, stroscio1993scanning}. A tunneling current is generated by scanning a sharp conducting tip across the sample surface with a nanometer-scale separation. STM has been extensively employed to study both the topography, as well as spectroscopic properties of moir\' {e} systems (top and bottom panels in Fig.~12a). In the topography mode, variations in the local density of carriers at AA and AB sites, provide a direct measurement of the size of the moir\' {e} unit cell, and therefore the twist angle~\cite{PhysRevLett.106.126802, PhysRevLett.109.126801, li2010observation, kerelsky2019maximized, choi2019electronic, jiang2019charge}. STM is also useful in capturing broken spatial symmetries that modify electron localisation leading to distinct density-of-states profiles. For instance, in the TDBG system, STM measurements have reported the breaking of three-fold rotational symmetry ($C_3$) within the non-dispersive bands while other lattice symmetries remain preserved~\cite{rubio2022moire}. The unidirectional stripe-like features seen at specific gate voltages within the flat moir\' {e} conduction band resemble the `nematic' ordering of electrons, seen previously in strongly correlated systems~\cite{avci2014magnetically, fernandes2014drives}. Interestingly, the nematic order was found to be independent of the applied displacement field, which breaks $C_2$ symmetry but preserves $C_3$ symmetry. Although the recent observation of anisotropic bulk transport properties near half-filling in MATBG also suggests a nematic phase~\cite{cao2021nematicity}, STM topography measurements provide direct evidence for such spatial ordering.\\

Additionally, the scanning tunneling spectroscopy (STS) technique is routinely employed to identify the many-body origin of various gapped phases and to quantify the characteristic topological invariants via the application of a magnetic field. Most significantly, the local density of states (LDOS) revealed through differential conductance (d$I$/d$V$) measurements demonstrate the emergence of vHSs at half-fillings in the AA-stacked region for different twist angles~\cite{kerelsky2019maximized, PhysRevLett.109.126801, li2010observation} (first plot in the bottom panel of Fig.~12a). The energy separation between these peaks reduces as the twist angle decreases, indicating extremely flat bands at the magic angle. Moreover, STS data in MATBG have detected Chern insulators and flavour symmetry breaking. The application of a strong perpendicular magnetic field leads to a sequence of topologically gapped Chern insulators at different integer fillings~\cite{nuckolls2020strongly, choi2021correlation} (second plot in the bottom panel in Fig.~12a). The calculated Chern numbers from the spectroscopic gaps are consistent with the $\mathcal{T}$ symmetry-breaking mechanism (discussed in Fig.~10). The STS data in MATBG at zero magnetic field, further revealed a cascade of transitions that shed light on the interplay between electron-electron interactions and quantum degeneracy (third plot in the bottom panel in Fig. 12a)~\cite{wong2020cascade}.~The splitting of the degenerate flat bands into Hubbard subbands with a spectral gap determined by the on-site Coulomb repulsion $U$ has been proposed to explain these observations. This process gives rise to nine minimum-energy Hubbard subbands at each integer filling, including the charge neutrality point ($\nu=0$), consistent with the isospin flavour symmetry breaking at partial band fillings. Such rearrangements of the low-energy excitations at  integer moir\'{e} band fillings, are also aligned with the isospin flavour reordering discussed in Fig.~8.\\

In the hole-doped superconducting phase of MATBG, point-contact spectroscopy and STS have been utilised to determine the superconducting gap directly~\cite{oh2021evidence}. The data suggests unconventional superconductivity with a V-shaped gap at low temperatures and a pseudo-gapped phase at higher temperatures and magnetic fields. U-shaped spectra are generally observed in conventional s-wave superconductors, while V-shaped spectra result from nodal superconducting gaps (fourth plot in the bottom panel in Fig. 12a). The observation of a large tunneling gap in MATBG exceeded the values expected from a BCS-type mechanism. Similar observations reported in MATTG also suggested an unconventional superconductivity~\cite{kim2022evidence}. The absence of these features when MATBG was aligned with hBN indicates the importance of the structural characteristics and the $C_2$ symmetry of unaligned MATBG in stabilising these ground states~\cite{oh2021evidence}. Although it was not possible to completely rule out a phonon-based pairing mechanism, the findings demonstrated a violation of BCS-type superconductivity providing essential constraints for developing an accurate theory for superconductivity in MATBG.\\

While STM can delve into the details of the atomic arrangement and band structure, such information about the ground states may not be sufficient to understand thermodynamic phase transitions as the moir\'{e} bands are partially filled. Any spontaneous process in the presence of broken symmetries is governed by thermodynamics. For instance, a system undergoes a phase transition if the change in free energy of the process is negative, making it thermodynamically stable. To this end, electronic compressibility measurements can be used to track changes in the chemical potential of the moir\'{e} system. Bulk compressibility, measured as a function of carrier density, maps the correlated phases seen previously in transport measurements (see Fig.~8). Additionally, the thermodynamic bandwidth can be directly estimated, giving insights into the flatness of the low-energy bands. Even on a local scale, electronic compressibility (expressed as d$\mu$/d$n$) can be measured using a scanning-probe SET (left panel in Fig.~12b). These experiments have revealed multiple features with negative compressibility regions identified as correlated insulators~\cite{zondiner2020cascade, pierce2021unconventional, PhysRevLett.123.046601}, symmetry-broken Chern insulators~\cite{pierce2021unconventional, xie2021fractional, yu2022correlated}, and CDW-like states~\cite{pierce2021unconventional, xie2021fractional} stabilised in different ranges of the magnetic fields. The isospin flavour symmetry breaking leads to the pinning of chemical potential at each integer filling (right panel in Fig. 12b). This is equivalent to a phase transition from a compressible (positive compressibility) to an incompressible/gapped (negative compressibility) state at the integer fillings. Compared to bulk transport measurements, the scanning SET technique was found to be superior in probing fragile electronic states such as fractional Chern insulators and CDW~\cite{pierce2021unconventional, xie2021fractional}. The technique has also been useful in estimating the entropy of the system, particularly in the context of the isospin Pomeranchuk effect discussed earlier~\cite{saito2021isospin, rozen2021entropic}.~Another recent SET experiment showcased a sequence of Chern insulators and multiple topological phase transitions at high magnetic fields~\cite{yu2022correlated}. The application of a strong out-of-plane magnetic field introduces Hofstadter subbands and triggers transitions among competing states with varying Chern numbers that are captured by the SET. Overall, scanning SET measurements have enriched our understanding of the thermodynamics governing the correlated ground states.\\

Next, we discuss one of the newest scanning techniques developed in recent times, namely the QTM~\cite{inbar2023quantum}. All of the present-day scanning methods probe the electronic wavefunctions at a single point in space at a time, effectively monitoring electrons as particles. Detecting the wave nature of electrons distributed over a large area has remained a  challenge. QTM was developed based on the quantum tunneling of electrons occurring via coherently interfering paths from the sample to the tip. The tool combines the working principles of atomic force microscope (AFM) and STM. Here, the key idea is to place a van der Waals material on an AFM tip creating a planar 2D junction ($\sim$~50 nm$-1\mu$m laterally) with the sample (another 2D layer of interest) kept underneath. QTM can be operated in two modes. The first mode allows continuous tuning of the twist angle between the 2D layer mounted on the tip and the 2D sample on the stage, leading to the possibility of probing electronic band structures of various multilayer twisted heterostructures (top left panel in Fig.~12c). In the second mode, a tunnel barrier is inserted between the two 2D layers. Unlike STM, which images electron wavefunctions in real space, this method of momentum-resolved tunneling probes electrons in the momentum space. The energy bands in TBG were imaged at room temperature by measuring the tunneling between a monolayer graphene and a TBG layer using WSe$_2$ as a tunnel barrier (bottom left and right panels in Fig.~12c). This new scanning probe is niche and complements STM, providing new opportunities for creating van der Waals interfaces with control over the twist angle, and accessing the energy dispersions of those systems. Implementing this technique at lower temperatures, where correlated physics comes into play, is expected to demonstrate its full set of capabilities in the years to come.\\

Finally, we discuss another scanning tool, the SOT, capable of imaging the domains spatially in a magnetic material. The working principle of this setup is different from that of STM, SET, and QTM, which measure the quantum tunneling of electrons. A SQUID, typically made of two Josephson Junctions, is extensively used as a magnetometer to detect tiny magnetic fields. SOT employs a Pb SQUID device on the apex of a sharp pipette, with a typical diameter of 200 nm~\cite{doi:10.1021/acs.nanolett.6b02841, uri2020mapping, tschirhart2021imaging} (top left panel in Fig.~12d). The SQUID is mounted on a quartz tuning fork and continuously scans the plane of TBG at a constant height of about 20-30 nm. The output signal is the SQUID's response to the lateral oscillations of the tip when an electrical excitation is applied to the quartz tube. The extreme sensitivity of SOT was first used to obtain tomographic images of the Landau levels in the quantum Hall state in MATBG, thereby providing a map of the twist angle disorder in the sample~\cite{uri2020mapping}. It was observed that even in presumably high-quality samples exhibiting correlated states and superconductivity, the twist angles locally vary to the tune of $\approx 0.1\degree$, establishing twist angle disorder as one of the most prominent defects in moir\'{e} systems.~In subsequent works, SOT was used to establish the nature of magnetism in MATBG~\cite{tschirhart2021imaging}. Although magnetism in MATBG was predicted to be orbital in nature, directly probing the magnetisation in moir\'{e} graphene systems has been tricky, particularly because of the low expected magnetisation density. The sensitivity of SOT was leveraged to image submicrometer-scale magnetic structures in MATBG, aligned with hBN. Magnetisation of several Bohr magnetons per charge carrier was observed (bottom left panel in Fig.~12d), confirming the presence of magnetism, primarily governed by the orbital degree of freedom. In a different study, a mosaic of microscopic domains was reported, with different Chern numbers that can remain irreversible upon changing the direction of the density sweep (right panel in Fig.~12d)~\cite{grover2022chern}. As a result, there is no hysteretic response, although the sample possesses local magnetic moments. These observations are consistent with the difficulties in reproducing AHE in bulk electrical transport, which requires a hysteresis in Hall resistance to confirm ferromagnetism. More recently, SOT was also employed to investigate other moir\'{e} systems such as the hetero-bilayer-TMDC system MoTe$_2$/WTe$_2$~\cite{tschirhart2023intrinsic}.

\begin{figure*}[tb!]
\includegraphics[width=0.8\textwidth]{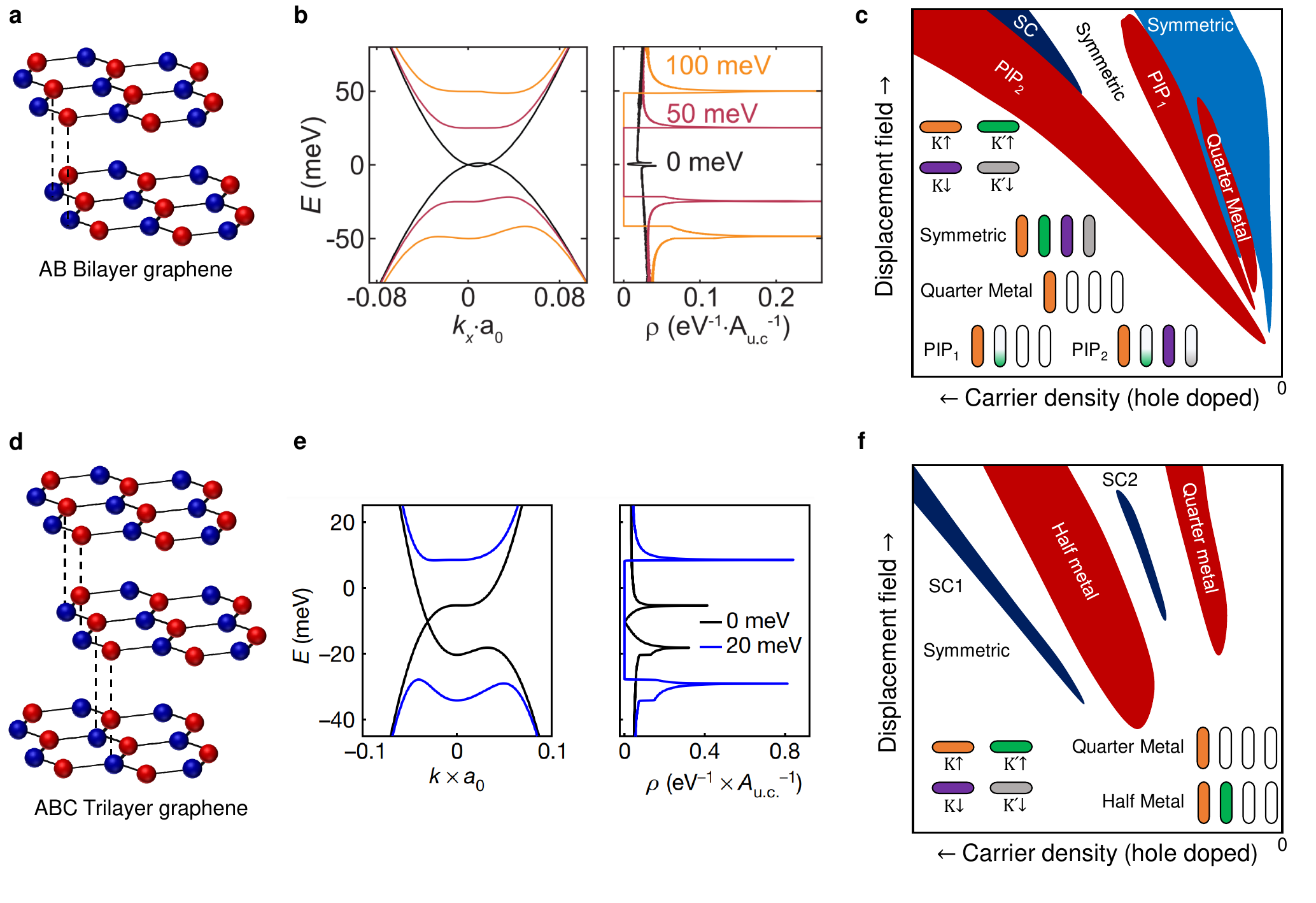}
\justify{\textbf{Fig.~13.~Phase diagram of Bernal bilayer and rhombohedral trilayer graphene.}~\textbf{a.}~Crystal structure of Bernal (AB) bilayer graphene. A and B sublattices are red and blue in colour.~\textbf{b.}~On the application of an electric field (50 or 100 meV), the bands become flatter with strongly diverging DOS. The band structure calculations are adapted with permission from ref.~\cite{zhou2022isospin}.~\textbf{c.}~Schematic compiling multiple phases observed in compressibility and resistance measurements. The isospin flavours are illustrated using four different colours. The phases are symmetric (isospin degeneracy 4), quarter metal (isospin degeneracy 1), and partially isospin polarised (PIP$_{1, 2}$). Superconductivity (SC) appears at the boundary of the symmetric and PIP$_2$ phase. The band structure of BLG is intricate, with multiple Fermi pockets due to trigonal warping. These are not explicitly identified in the simple schematic presented here; we refer the readers to ref.~\cite{zhou2022isospin} for a more detailed breakdown of the phase space.~\textbf{d.}~Crystal structure of rhombohedral (ABC) trilayer graphene.~\textbf{e.}~The band structure calculations adapted with permission from ref.~\cite{zhou2021half, zhou2021superconductivity} show a gap opening at the CNP followed by the formation of flat bands and a large DOS when a perpendicular electric field of 20 meV is applied.~\textbf{f.}~Quarter metal and half metal appear when the carrier density and electric field are varied. Two superconducting domes, SC1 and SC2, appear in two different regimes in the phase space.}
\end{figure*}

\section{Emergent phases in graphene without moir\'{e} superlattices}

Our discussions so far have been limited to moir\'{e} graphene layers exhibiting flat bands due to a slight mutual rotation between them, leading to a diverging DOS that amplifies Coulomb interactions. In this section, we discuss naturally occurring bilayer and trilayer graphene systems in their pristine form, without any artificial moir\'{e} pattern. As mentioned earlier, the introduction of an out-of-plane electric field transforms the low-energy parabolic bands of Bernal BLG into flatter bands, accompanied by a gap opening at the CNP~\cite{mccann2013electronic, zhou2022isospin} (Fig.~13a-b). The electronic structure of Bernal BLG mimics the larger class of rhombohedral multilayer graphitic systems and therefore shares similarities with ABC trilayer graphene. Calculations of the single-particle band structure in both these materials reveal a large DOS near the band edge that replicates the flat band condition discussed for MATBG~\cite{zhou2021half, zhou2021superconductivity} (see Fig.~13b and 13e). This distinct similarity prompted renewed interest in Bernal bilayer and ABC trilayer graphene, and true to the expectations, both these systems have shown remarkable phase diagrams, encompassing superconductivity and isospin magnetism.\\

The recent compressibility and magnetotransport study by Zhou et al.~\cite{zhou2022isospin} demonstrated a rich phase diagram in Bernal BLG as a function of the displacement field, carrier density, and magnetic fields. When the carrier density and electric field were varied simultaneously, the system was found to undergo multiple phase transitions, as indicated in Fig.~13c. These transitions, captured by magneto-oscillation measurements, were attributed to changes in the spin/valley isospin degeneracy of the Fermi surface. At lower densities near the CNP, the system can favour only one of the four flavours (two spins, two valleys), resulting in a state defined as the `quarter metal'. These phases exhibit ferromagnetic properties, which are consistent with the Stoner model, where the interplay between a high DOS and strong interactions favours ferromagnetism. However, spin/valley polarised quarter metal differs from conventional ferromagnets, where up and down spins are present in unequal proportions. Furthermore, when the device was cooled to temperatures less than 30 mK and a small in-plane magnetic field was applied in addition to the large out-of-plane electric field, the system became superconducting. Interestingly, superconductivity was not observed without an in-plane magnetic field. These observations suggest the possibility of a spin-triplet pairing mechanism rather than spin-singlet BCS-type superconductivity, a signature that was also seen in MATTG discussed earlier.\\

Transport properties of rhombohedral trilayer graphene (ABC-TLG) (Fig.~13d) under electric and magnetic fields, also showed a very similar phase diagram. A cascade of phase transitions separated by negative compressibility was observed by varying the carrier density and electric field on the hole-doped sides~\cite{zhou2021half, zhou2021superconductivity}. Detailed analysis revealed that on adding carriers, the degeneracy of the isospin flavours in the system (two spins, two valleys) governs these phases, referred to as `half metal' and `quarter metal' (Fig.~13f). At an intermediate density, when electrons with the same spin occupy $K$ and $K^\prime$ valleys, the system is called a half metal. At lower densities, the system also exhibits a quarter metallic state. Both these phases are isospin-polarised and exhibit ferromagnetism. Two distinct superconducting regions, SC1 and SC2, were found in two different regimes of the phase space. SC1, with a transition temperature below 106 mK emerges from the symmetric phase at a higher density where all four isospin flavours are equally filled. SC2 appears from the half-metallic phase below the lowest temperature in their measurement setup (20 mK). It was observed that SC1 disappears at an in-plane magnetic field of 300 mT, whereas SC2 survived up to 1 Tesla, which is an order of magnitude higher than its Pauli limit. The magnetic field dependence suggests that SC1, originating from a symmetric phase (isospin unpolarised), and SC2, stemming from a half-metallic phase (spin-polarised, valley unpolarised), are governed by spin-singlet and spin-triplet pairing mechanisms, respectively. It is to be noted that previously, the observation of ferromagnetism~\cite{chen2020tunable} and signatures of superconductivity~\cite{chen2019signatures} (with a finite residual resistance) in ABC-TLG aligned with hBN were explained, considering the interplay between electron-electron interactions within the flat band and the electric field tunable bandwidth. However, Zhou et al. found no significant difference in the observed phases when a moir\'{e} superlattice was introduced using ABC-TLG/hBN. Therefore, it is now questionable whether a superlattice potential is essential to engineer correlated phases in ABC-TLG or if they can emerge naturally in the pristine regime of any conventional graphene system featuring diverging D. Nevertheless, the presence of exotic phases in naturally occurring AB-BLG and ABC-TLG offers new material avenues, free from the complexities typically associated with twist angles. Future experiments are anticipated to unveil the origins of these phases and provide more insights into their striking similarity with the general class of graphene moir\'{e} systems.

\begin{figure*}[tb!]
\centering
\includegraphics[width=0.9\textwidth]{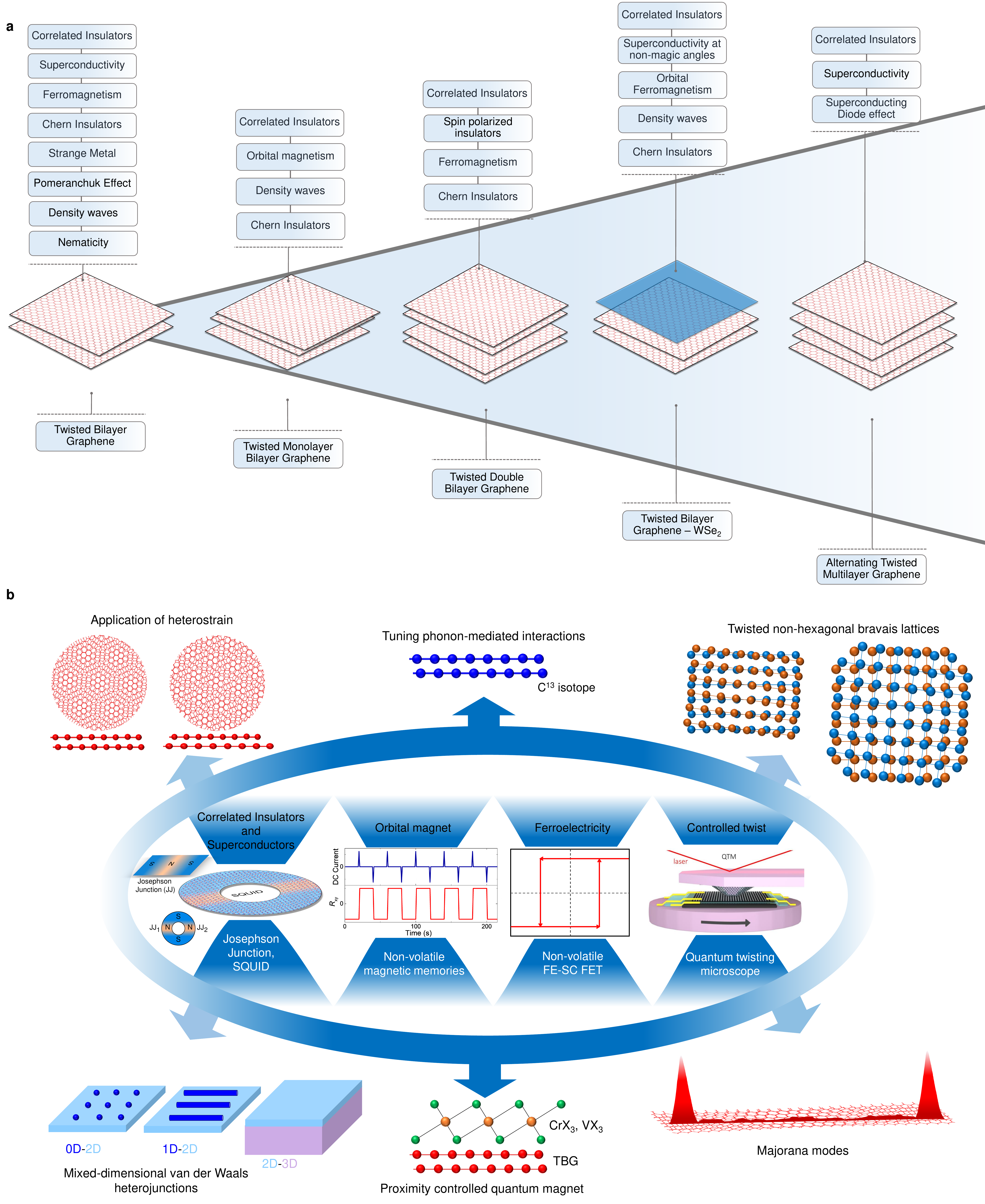}
\justify{\textbf{Fig.~14.~Outlook and future of moir\'{e} materials}~\textbf{a.}~Schematic capturing the expansion of the family of twisted graphene systems with time. The correlated phases observed in each of these systems are listed at the top.~\textbf{b.}~The inner circle shows the emergent phases in TBG and some of their potential applications, which are already being developed. Many more possibilities are on offer, as listed outside the circle.}
\end{figure*}

\section{Outlook}

The family of moir\'{e} materials has been expanding rapidly with novel fabrication strategies enabling the creation of complex stacks that combine 2D layers. We have reviewed, at least partially, the immensely rich and complex collection of quantum phases that have been found in graphene-based moir\'{e} materials. It is evident that the unprecedented tunability with the introduction of the twist angle between different layers as a new degree of freedom has opened this exciting new world. As the field develops, addressing outstanding experimental and theoretical challenges must go hand-in-hand with exploring the potential to utilise moir\'{e} heterostructures in disruptive technological applications. On the fundamental front, the complexity of the phase diagram and its sensitivity to small changes in external parameters indicate a close competition between different phases of matter in moir\'{e} materials, and the phase usually detected in the laboratory is the one that minimises the free energy. While the most common experimental tools for mapping the phase diagram utilise global transport properties, these efforts are beginning to be augmented with thermodynamic and spectroscopic probes, such as tunneling microscopy~\cite{nuckolls2020strongly, kerelsky2019maximized, kim2022evidence, choi2019electronic, jiang2019charge}, spatially resolved magnetic mapping~\cite{uri2020mapping, tschirhart2021imaging, grover2022chern}, THz-spectroscopy, and scanning near-field
optical microscopy~\cite{hesp2021nano, hesp2021observation, halbertal2022unconventional} for transient quasi-particle kinetics, and others. In parallel, well-organised theoretical efforts need to look for a global `order parameter’, if any, that can connect the apparently diverse experimental results with a common thread. A consolidated theoretical framework is essential also to predict new moir\'{e} structures with nontrivial properties, e.g., fractional Chern insulators, unconventional superconductivity, and so on that experimentalists may attempt realising in laboratories.\\

Below we discuss the opportunities and challenges associated with fabricating, measuring, and potentially utilising moir\'{e} materials as building blocks for technological applications. In Fig.~14a, we illustrate the evolution of various complex moir\'{e} heterostructures over time. Despite differences in the number and types of layers, dielectric environment, twist angle, and symmetries, these structures feature commonalities in terms of their band structures, and responses to external stimuli such as electromagnetic field, temperature, and pressure. Stabilising emergent properties by carefully engineering the band structure of moir\'{e} systems would be an outstanding development. A unique approach to tuning the dielectric environment can be achieved by using different materials in proximity to the twisted graphene layers. Dielectric engineering using a broad range of 2D materials, particularly TMDCs and magnetic insulators such as chromium trihalides, holds promise as a future research direction~\cite{PhysRevMaterials.2.024004, jiang2018controlling, doi:10.1021/acs.nanolett.2c02931}.\\

Another exciting possibility involves qualitatively tuning the band structure through the application of a mechanical strain. Despite the much-revered band tunability achieved by the unique experimental knob, namely the twist angle, the flat bands and correlated electronic states in moir\'{e} materials are fragile and can be altered by even small structural deformations~\cite{kazmierczak2021strain, wang2023unusual, doi:10.1021/acsanm.0c03230}. While intrinsic intralayer atomic lattice reconstruction processes have been noted to introduce intralayer strain, extrinsic uniaxial heterostrain is also believed to strongly manipulate the symmetries and ground states in these artificial materials. The latter is expected to primarily occur when there is an in-plane deformation or stretching of one layer with respect to another, and is predicted to change the bands and alter the moir\'{e} potential in dramatically influential ways in comparison to homostrain~\cite{PhysRevB.98.235402, PhysRevLett.127.126405}, and, therefore, may enable the discovery of new classes of emergent behaviour, beyond those observed thus far in the absence of heterostrain-control. While unintentional, inherent strain effects have recently been inferred from non-linear Hall measurements in twisted double bilayer graphene samples~\cite{sinha2022berry, chakraborty2022nonlinear} and also seen as atomic reconstructions in marginally TBG samples~\cite{yoo2019atomic}, consolidated efforts to tune the strain-environment in moir\'{e} heterostructures would form a promising research direction.\\

Our discussions have so far pertained to electron-electron interactions without focusing on the contributions of phonons since many of these correlated states appear at ultra-low temperatures where phonon contributions are expected to be minimal. However, electron-phonon coupling can enter as a leading mechanism for superconductivity, which is still a debated topic. Therefore, studying the role of phonons in these systems can prove to be instructive. The frequency of different phonon modes generally depends on the effective mass of the atoms. Tuning the overall band structure by making TBG samples with C$^{13}$ isotope instead of C$^{12}$ can provide insights into the strength of phonon interactions.\\

Research beyond twisted graphene involves studying new materials, such as superlattices composed of triangular and rectangular lattices, which would significantly extend the moir\'{e} family. Twisted bilayers of TMDC and twisted GeSe layers are ideal candidates to generalise the effect of moir\'{e} potentials. Bilayer GeSe and GeS can host strongly collective phenomena like Luttinger liquid, Mott insulators, and charge-ordered states~\cite{kennes2020one, wang2022one}. Another open research direction is exploring whether the large SOC can stabilise correlated phases in these systems. Combining 2D layers with their 0D, 1D, and 3D counterparts is yet another avenue to create a new class of materials. Mixed-dimensional heterostructures, with a 2D layer in proximity to their 3D, 1D, and 0D components, can lead to unique material platforms that can influence the overall band structure. Moreover, a finite twist angle between the individual components with different dimensions adds an extra knob to control the properties of the hybridised system. Preliminary studies on moir\'{e} lattices on graphitic thin films promise more surprises~\cite{waters2023mixed, mullan2023mixing}.~Recently, superconductivity has been reported in quasiperiodic TTG, where the three graphene layers are rotated by two different angles~\cite{uri2023superconductivity}.~Such a structure is composed of two moir\'{e} patterns which are incommensurate with respect to each other. Varying the number of layers and twist angle can give rise to an infinite number of quasicrystals which are anticipated to be fascinating platforms for uncovering electronic phases.\\

Complex challenges accompany outstanding opportunities. The need for a finite twist angle in graphene-based moir\'{e} systems has been a double-edged sword. Fabricating samples without twist angle disorder has been extremely difficult, with studies showing that no two samples have the same structural and electronic properties. The quality of the devices degrades due to unwanted twist-angle inhomogeneity, strain, and static disorder introduced while fabricating heterostructures. Surprisingly, even two TBG samples with the same twist angle, as estimated from transport measurements, have exhibited widely different properties, indicating that there is an unknown set of parameters governing the underlying mechanism of the correlated states. A better fabrication protocol with precise control over limiting factors, such as twist angle, strain, and disorder, over the micrometer scales will be necessary. Several important fabrication strategies have been adopted in recent years to improve the quality of the moir\'{e} samples. Using a graphite gate can lead to homogeneous doping of charge carriers compared to conventional metal gates such as gold and doped silicon, which may produce non-uniform doping due to the presence of grains, charge puddles, or trapped states~\cite{zibrov2017tunable, spanton2018observation}. The transport data stands as a testimony to this hypothesis, with some of the best-correlated signatures being demonstrated in graphite-gated samples. In another advance over the `tear and stack' method, pre-cutting the graphene flake into two halves using a sharp tip or a laser was found to provide better angle homogeneity~\cite{saito2020independent, park2021flavour}.~This method of `cut and stack' avoids the tearing of graphene, which causes a significant strain on graphene layers, leading to the relaxation of their orientation to a random direction instead of the desired one. Developing more such fabrication strategies is an absolute need for reliable devices.\\

Limitations in large-scale production and reproducibility of the observed phases have restricted the current applicability of moir\'{e} materials to technologies. However, they hold promise as ideal candidates for many future device applications. Many of the current state-of-the-art quantum devices rely on various semiconductor heterostructures that are intricately fabricated to create suitable junctions. Such structures inherently suffer from interface defects, lattice mismatches, and structural distortions. One of the ways to combat these issues is to design a monolithic platform where a circuit encompassing various components can be designed using the same material. Such an effort using conventional platforms such as silicon has not been possible to date. Moir\'{e} systems are ideal candidates to bridge this gap. In particular, tunability between superconducting and insulating phases in moir\'{e} materials via simple electrical voltages naturally provides a new testbed for Josephson Junctions (JJ), superconducting diodes, SQUIDs, and other superconducting quantum devices~\cite{rodan2021highly, de2021gate, portoles2022tunable, diez2023symmetry}.~The ability to electrically tune the same material to be a superconductor and an insulator offers new pathways to design JJs, consisting of two superconducting leads connected via a weak link. JJ are important elements with applications in superconducting qubits, bolometers, and low-temperature integrated circuits. An extension of JJs is the SQUID, which can be used as sensors to detect tiny magnetic fields. Recent reports have demonstrated proof-of-concept devices using MATBG and MATTG, which include Josephson Junctions~\cite{rodan2021highly, de2021gate}, SQUIDs~\cite{portoles2022tunable}, and superconducting diodes~\cite{diez2023symmetry, lin2022zero}. Electrically tunable orbital magnetism also holds promise for non-volatile memory devices~\cite{serlin2020intrinsic, polshyn2020electrical}. Additionally, recent demonstrations of ferroelectricity, along with stable switching between the superconducting, metallic, and correlated insulator states of MATBG, promise new kinds of switchable moir\'{e} electronic systems~\cite{klein2023electrical}.~The recent theoretical proposal of Majorana modes in spin-orbit coupled TBG also forms an interesting direction toward topological qubits~\cite{PhysRevB.105.L081405}.\\ 

In summary, all the surprises moir\'{e} materials have featured to date promise exciting opportunities for new scientific discoveries and technological advancements. The intricate nature of these materials demands sophisticated experimental techniques and theoretical models. Better synthesis and characterisation protocols are the most essential requirements at this stage. It is anticipated that these obstructions can be encountered at a fast pace through interdisciplinary collaborations and the collective effort of several research groups. We believe this is only the beginning of a new field that can continue to serve as a promising platform for exploring unconventional quantum phenomena with vast potential applications.\\

\section*{Acknowledgements}
U.C. acknowledges funding from SERB via SPG/2020/000164 and WEA/2021/000005.

\bibliographystyle{naturemag}
\bibliography{References}

\end{document}